\documentclass[manuscript,screen]{acmart}
\usepackage{subcaption}
\usepackage{multirow}
\usepackage{makecell}  
\usepackage{graphicx}  
\renewcommand\footnotetextcopyrightpermission[1]{}
\setcopyright{none}
\usepackage{physics}
\AtBeginDocument{%
  }

\begin{document}

\title{AC/DC: Automated Compilation for Dynamic Circuits}

\author{Siyuan Niu}
\email{siyuanniu@lbl.gov, siyuan.niu@ucf.edu}
\orcid{0000-0003-4683-381X}
\affiliation{%
  \institution{Computational Research Division, Lawrence Berkeley National Laboratory}
  \city{Berkeley}
  \state{California}
  \country{USA}
}

\author{Efekan K\"okc\"u }
\orcid{0000-0002-7323-7274}
\affiliation{%
  \institution{Computational Research Division, Lawrence Berkeley National Laboratory}
  \city{Berkeley}
  \state{California}
  \country{USA}
}

\author{Anupam Mitra}
\orcid{0009-0004-6711-2477}
\affiliation{%
  \institution{Computational Research Division, Lawrence Berkeley National Laboratory}
  \city{Berkeley}
  \state{California}
  \country{USA}
}

\author{Aaron Szasz}
\orcid{0000-0002-1127-2111}
\affiliation{%
  \institution{Computational Research Division, Lawrence Berkeley National Laboratory}
  \city{Berkeley}
  \state{California}
  \country{USA}
}

\author{Akel Hashim}
\orcid{0000-0002-8611-0125}
\affiliation{%
  \institution{Quantum Nanoelectronics Laboratory, Department of Physics,
University of California, Berkeley and Computational Research Division, Lawrence Berkeley National Laboratory}
  \city{Berkeley}
  \state{California}
  \country{USA}
}

\author{Justin Kalloor}
\orcid{0000-0001-6749-5501}
\affiliation{%
  \institution{Department of Electrical Engineering and Computer Science, University of California, Berkeley}
  \city{Berkeley}
  \state{California}
  \country{USA}
}

\author{Wibe Albert de Jong}
\orcid{0000-0002-7114-8315}
\affiliation{%
  \institution{Computational Research Division, Lawrence Berkeley National Laboratory}
  \city{Berkeley}
  \state{California}
  \country{USA}
}

\author{Costin Iancu}
\orcid{0000-0001-7845-2427}
\affiliation{%
  \institution{Computational Research Division, Lawrence Berkeley National Laboratory}
  \city{Berkeley}
  \state{California}
  \country{USA}
}

\author{Ed Younis}
\orcid{0000-0002-1306-1860}
\email{edyounis@lbl.gov}
\affiliation{%
  \institution{Computational Research Division, Lawrence Berkeley National Laboratory}
  \city{Berkeley}
  \state{California}
  \country{USA}
}

\renewcommand{\shortauthors}{Niu et al.}
\newcommand{\SY}[1]{{\color{red}{#1}}}
\newcommand{\EK}[1]{{\color{magenta}{#1}}}
\begin{abstract}
Dynamic quantum circuits incorporate mid-circuit measurements (MCMs) and feed-forward operations are crucial for manipulating quantum information. They have been broadly used in quantum error correction and quantum teleportation. Recently, they are utilized to prepare certain states and long-range entangling gates as well as reduce resource overhead in quantum algorithms.
In this paper, we present AC/DC, a novel \underline{A}utomated \underline{C}ompilation framework for generating \underline{D}ynamic quantum \underline{C}ircuits that prepare any unitary operators or states, leveraging numerical optimization-based circuit synthesis methods.
The first contribution is introducing optimization objective functions incorporating MCMs and feed-forward operations. The second contribution is embedding these into a popular open-source quantum circuit synthesis framework. We demonstrate generating dynamic circuits for long-range entangling gates, circuit optimization, lattice simulations, and state preparation, with validation through simulation and quantum hardware. Furthermore, we perform a noise analysis to assess the impact of MCM and gate errors, identifying scenarios where dynamic circuits provide significant benefits. The dynamic circuits generated by our framework show substantial improvements in reducing circuit depth and, in some cases, the number of gates. To our knowledge, this is the first practical procedure to generate dynamic quantum circuits, paving the way for enhanced circuit generation and optimization methods for near-term quantum computers.

\end{abstract}




\maketitle

\section{Introduction}
\label{sec:introduction}

Dynamic or adaptive quantum circuits incorporate non-unitary
mid-circuit measurements (MCMs) and classical feed-forward
control~\cite{corcoles2021exploiting,pino2021demonstration,graham2023midcircuit}. These
type of circuits are at the core of the implementation of any Quantum
Error Correction (QEC) code. Meanwhile, on the road to Fault Tolerant
Quantum Computing (FTQC), they have been used to prepare quantum states, such
as matrix-product states, and long-range entanglement gates in
constant depth~\cite{baumer2024efficient, hashim2024efficient,
  smith2024constant, smith2024constant, baumer2024measurement,
  buhrman2023state, smith2023deterministic,zhang2024characterizing}. In these studies,
circuits are derived manually using quantum-teleportation
protocols. Other works show how to reduce resource requirements for
Quantum Fourier Transformation~\cite{baumer2024quantum} and Quantum
Phase Estimation~\cite{corcoles2021exploiting}. Dynamic circuits also  enable scaling up distributed quantum computing systems, by sharing quantum states across multiple quantum
processors~\cite{vazquez2024scaling}.  Additionally, parameterized
dynamic circuits have been shown not to suffer from the barren plateau
problem~\cite{deshpande2024dynamic}.

Despite their great potential, generally applicable methods for
generating arbitrary dynamic circuits do not yet exist.  Most methods
rely heavily on Bell state teleportation, stabilizer formalism, and
group theory, which require extensive manual
fine-tuning~\cite{foss2023experimental, iqbal2024topological}. To our
knowledge, the single exception is a study that introduces a
variational method for using dynamic circuits to prepare generic
quantum states~\cite{alam2024learning}. Providing a scalable generic
method will facilitate advances in all the use cases introduced so
far, as well as novel circuit optimization methods. Furthermore, MCM
with feed-forward implementations have been only very recently
demonstrated on hardware with limited examples. Their dynamics are
still not well understood, and no benchmarks are available to test
these capabilities at the algorithm level. An automated generation
method can be easily  used to build algorithm level hardware
benchmarking suites, similar to~\cite{tomesh2022supermarq, lubinski2023application}.

\begin{figure}[t]
    \includegraphics[scale=0.8]{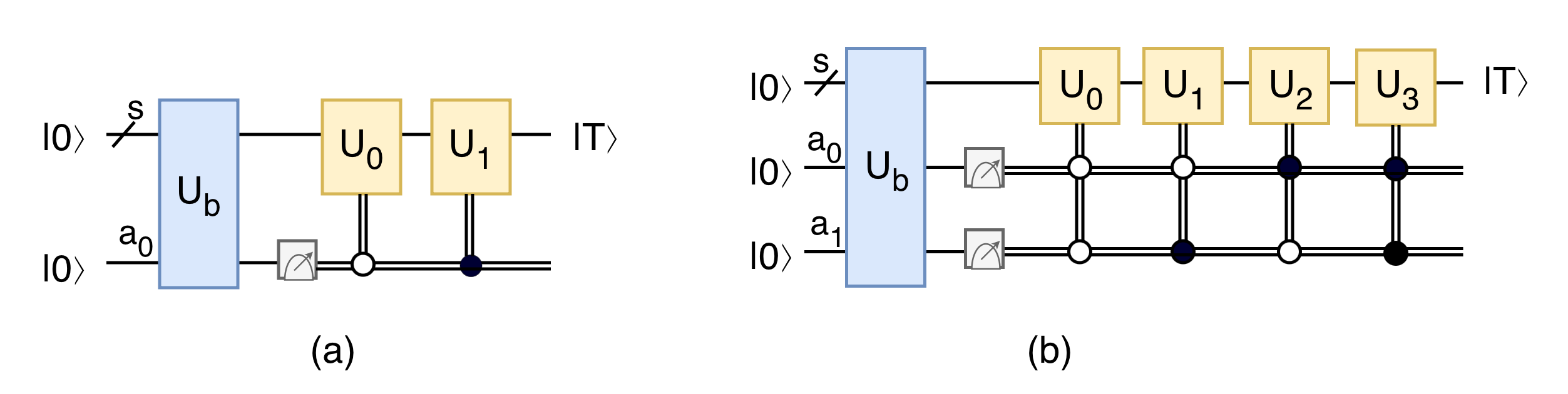}
    \caption{
    Dynamic circuit protocol for preparing a target state $\ket{T}$ using (a) 1 ancilla and (2) 2 ancillas with simultaneous measurement in one cycle. $s$ denotes the system qubits and $a_i$ denotes the ancilla qubits. It is easy to extend the protocol to arbitrary number of ancillas. The goal is to determine $U_b$ and $U_i$ for $i=0, \dots, 2^{a-1}$. }
    %
    \label{fig:protocol-state}
\end{figure}

In this work, we employ unitary synthesis methods based on
parameterized circuit representations and numerical optimization. To
guide generation, this class of methods~\cite{cincio2021machine,davis2020towards,rakyta2022approaching,rakyta2022efficient,madden2022best} uses
objective functions based on Quantum Information Science process
distance metrics, such as Hilbert-Schmidt or Frobenius. As they can
manipulate only unitary processes, the first challenge to
generalization is deriving objective functions that can incorporate
non-unitary MCM and classical feed-forward operation. To the best of our knowledge, this is the first one to propose a novel objective function for preparing arbitrary unitaries using dynamic circuits with mid-circuit measurement and feed-forward operations, described in Section~\ref{sec:unitary}. We also propose an alternative cost function different from the state-of-the art for state preparation described in Section~\ref{sec:state}. These objective functions have been derived to
combine operational semantics with computational
tractability.

The circuit transformations required for practical demonstrations are
presented in Figure~\ref{fig:protocol-state} and Figure~\ref{fig:protocol-unitary}.  To perform these, we extend a
popular quantum circuit synthesis framework BQSKit~\cite{osti_1785933} with novel
generation algorithms, and showcase several use cases. For unitary preparation, we demonstrate the creation of long-range
generic entangled gates and multi-qubit gates, and propose scaling to large
circuits using a circuit partitioning method. Additionally, we
demonstrate a physics application of dynamic circuits for lattice
simulation. For state preparation, we demonstrate generation of GHZ, W, and Dicke states. Besides introducing new circuit optimization methods, one
useful outcome of this effort is the ability to generate algorithm
level benchmarks for hardware validation and evaluation.

We validate the approach on simulator and hardware and perform noise modeling to
determine when dynamic circuits can be beneficial when compared to
equivalent unitary circuits, based on different error rates of
mid-circuit measurements and the two-qubit gates used. The results indicate that dynamic circuits offer new opportunities to explore trade-offs in the circuit generation process, highlighting their potential to enhance the performance of quantum algorithms.

Our major contributions can be summarized as follows:

\begin{itemize}

    \item We propose an optimization objective function for preparing arbitrary unitary operators that incorporates mid-circuit measurements and feed-forward operations. 
    
    \item We introduce AC/DC, a novel \underline{A}utomated \underline{C}ompilation framework for generating \underline{D}ynamic \underline{C}ircuits for arbitrary unitaries and states that integrates the objective functions and circuit synthesis process. 

    \item  We present customized dynamic circuit structures with distinct measurement scenarios and their corresponding cost functions, providing new design opportunities for dynamic circuits.

    \item We utilize our AC/DC framework to prepare multiple unitaries, states, and circuits for lattice simulation, all designed with short circuit depths, and we evaluate them via simulation and quantum hardware.

\end{itemize}

\section{Background}
\label{sec:background}

In this section, we first introduce the concept of dynamic circuits, their applications, and the design challenges on both the hardware and software sides. Then, we delve into quantum circuit synthesis, with a particular focus on methods based on numerical optimizations.

\subsection{Dynamic Circuits}
Most practical demonstrations of algorithms on quantum hardware use
digital (unitary) gates with measurements only at the end of the execution: 
all gates are applied without dependence on real-time data
produced during execution. In contrast, a dynamic circuit incorporates
real-time information, taking classical processing into account during
execution. This approach allows for the measurement of quantum states in the middle of the circuit and the use of feed-forward operations to pass readout information from the classical register back to the quantum circuit. Depending on the classical readout
information, different quantum operations are then applied.

Dynamic circuits can be illustrated as shown in Figure~\ref{fig:protocol-state}. In general, there are $s$ system qubits and $a$ ancilla qubits. In the circuit, a unitary transformation $U_b$ is first applied to all qubits, followed by the measurement of the ancilla qubits. As we will demonstrate in Section~\ref{sec:measure}, the placement of these measurements will vary. Depending on the measurement outcome $i = 0,1,\dots,2^a-1$, the feed-forward operation (represented by two lines) transmits the measurement outcome to the classical control system. Based on this outcome, different unitary $U_i$ will be applied to the system qubits. 
After the measurement of ancilla qubits and feed-forward operation, we can reset ancilla qubits to state $\ket{0}$ and reuse them again~\cite{decross2023qubit, niu2024effective}.

Besides their application in QEC, dynamic circuits can be used for
program resource optimization to improve performance. Here, performance
can be viewed as decreasing runtime or increasing execution
fidelity. As execution fidelity is directly translated into QEC runtime
overhead, these uses are valid in the NISQ and FTQC regimes.  Due to the recent availability of MCM hardware implementations, the trade-offs in code generation are poorly understood. It is necessary to combine gate count and depth impact criteria with MCM error rates and real-time performance, as well as consider the impact of adding the necessary ancilla qubits. Methods to asses the impact of resource optimization on performance are readily available in the literature~\cite{beverland2022assessing, fellous2023optimizing,lubinski2024optimization}.

In many hardware platforms, measurements are much slower and exhibit much higher
error rates than unitary gates. For example, on the
available IBM superconducting quantum hardware, measurement typically
takes around 1 $\mu$s with an error rate of approximately $1\%$. Gates
execution is on the $n$s scale, with error rates in the 0.1\% range
for two qubit gates. Besides measurement latency, code generation
needs to take into account the latency of real-time communication
required for the feed-forward loop: at best, this can take 100s of nanoseconds 
\cite{hashim2024efficient, fruitwala2024distributed}. Moreover, MCMs can lead to 
errors on unmeasured spectator qubits, such as measurement-induced dephasing 
\cite{hashim2024efficient}. These metrics will certainly improve, but the fact will remain that there is a trade-off between the errors and extended duration introduced by mid-circuit measurements and feed-forward loops, and the reduced circuit depth achievable with dynamic circuits. Additionally, the trade-off between the possible circuit depth reduction and the increase in width due to the inclusion of ancilla qubits is also poorly understood.

For the future success of quantum computing, all these interactions need to be studied. In this endeavor, an automated generation strategy for dynamic circuits is mandatory.

\subsection{Quantum Circuit Synthesis}

Synthesis is the process of translating a high-level description of a
quantum program into an executable circuit on quantum
hardware. Digital quantum gates are unitary operators/matrices, and a
quantum program can be represented as a large unitary matrix if no
non-unitary processes are involved. The synthesis process then becomes
decomposing the large unitary matrix into the hardware native gate
sets, which consist of single-qubit gates in $U(2)$ and two-qubit
gates in $U(4)$.

Many synthesis methods~\cite{cincio2021machine,davis2020towards,rakyta2022approaching,rakyta2022efficient,madden2022best, weiden2022wide,di2016parallelizing,younis2021qfast,xu2023synthesizing,shende2005synthesis} exist, all exhibiting a trade
off between solution quality and time to solution. Probably the most
scalable, but resulting in exponential circuit depth is the Quantum
Shannon Decomposition~\cite{shende2005synthesis}. At the other end of
the spectrum are the numerical optimization based methods~\cite{davis2020towards, younis2021qfast,xu2023synthesizing}. 

Due to good circuit quality, in this work we target directly numerical optimization based synthesis.
Given  a target unitary matrix $U_T$,
these synthesis algorithms will generate a unitary matrix $U_S$, such
as
\begin{equation}
    \|U_T - U_S\|_{HS} < \epsilon,
\end{equation}
where $\|\cdot\|_{HS}$ is in this example the Hilbert-Schmidt
norm. Different algorithms use different norms, but all need to have
operational semantics and a low computational overhead.

Intuitively,
the synthesis algorithm walks a space of implementations, while
maintaining a small $\epsilon$ and trying to improve program
performance. At each step, an {\it instantiation} operation is
performed: this tries to instantiate circuit parameters to minimize the
distance to target. For example, if we want to use the parameterized
circuit with unitary $U_S(\theta)$ to approximate a target unitary
$U_T$, instantiation solves the problem of
\begin{equation}
    \arg\min_{\theta} \|U_T - U_{S}(\theta)\|_{HS}
\end{equation}

Since depth and/or gate count are good proxies
for the likelihood of a program to successfully execute on hardware,
these metrics are often used to guide the space walk to a solution.

We take a very practical approach to building the compilation
framework able to generate dynamic circuits. We first develop
objective functions for instantiation, which are able to combine unitary and
non-unitary behavior. As a starting point we use the Hilbert-Schmidt
distance as it has already been well studied. We then modify the
BQSKit synthesis framework to incorporate these objectives in its synthesis workflows. This integration gives us several advantages.

First, we  leverage
an existing topology aware optimal algorithm called QSearch~\cite{davis2020towards} which uses an $A^*$ based search algorithm. Previous studies~\cite{liu2023tackling,weiden2022wide} indicate that topology
aware synthesis generate better quality circuits than traditional compilers (e.g. Qiskit~\cite{qiskit2024} or Tket~\cite{sivarajah2020t}) using  available good quality qubit mapping algorithms, such as Sabre~\cite{li2019tackling}.

Second, BQSKit uses a divide-and-conquer strategy to ensure
scalability with circuit size. A circuit is partitioned~\cite{wu2020qgo} in
small blocks, which are directly re-synthesized using one of the
available algorithms. These blocks are then stitched together in the
solution circuit, with verification methods~\cite{patel2022quest,osti_1785933} to
ensure correctness of the transformation.

\section{Unitary preparation}
~\label{sec:unitary}

\begin{figure}[t]
    \includegraphics[scale=0.8]{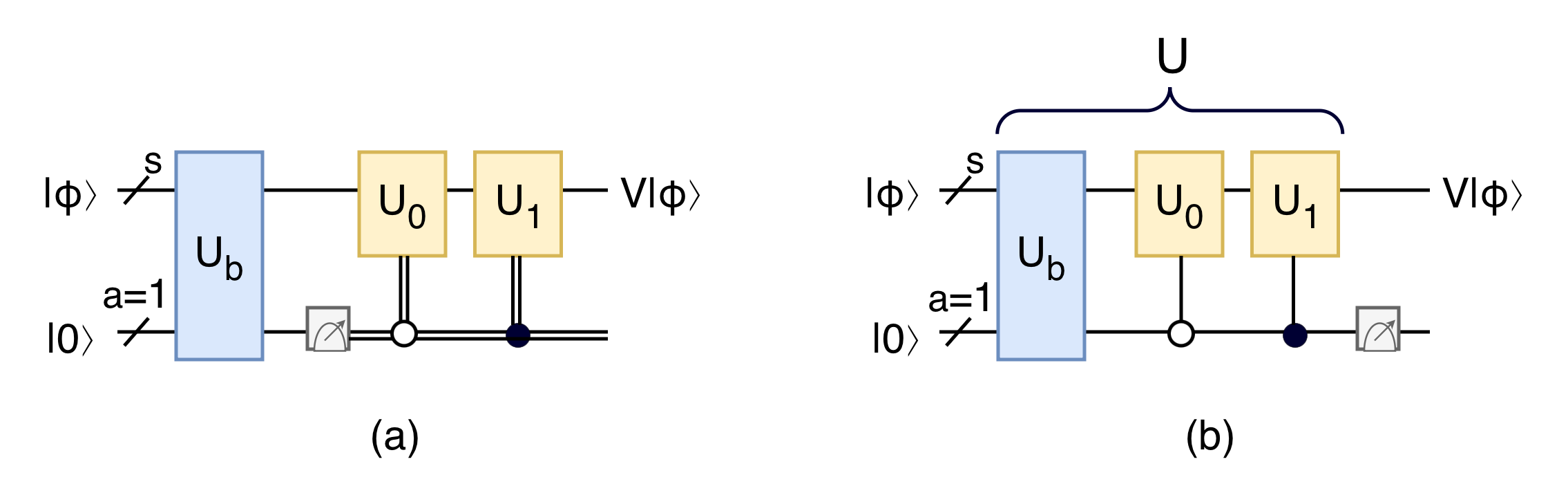}
    \caption{
    Panel (a) is the dynamic circuit protocol for preparing a target unitary $V$. $s$ denotes the system qubits, $a$ denotes the ancilla qubits, where we take it to be equal to $1$ for simplicity. We can extend the protocol to arbitrary number of ancillas. The goal is to determine $U_b$ and $U_i$ for $i=0, \dots, 2^{a-1}$. In panel (b), we illustrate how we can defer the measurement. This allows us to derive the unitary preparation cost functions $C_{\text{dyn1}}$ and $C_{\text{dyn2}}$ in Section \ref{sec:unitary}. 
    %
    }\label{fig:protocol-unitary}
\end{figure}

The protocol for unitary preparation using dynamic circuits is a generalization of the single ancilla qubit example in
Figure~\ref{fig:protocol-unitary} (a), to an arbitrary number
$a$ of ancilla qubits. In this section, we will use the simultaneous measurement of multiple ancilla qubits as an example to demonstrate how to integrate the measurement of the ancillas into known cost functions based on the Frobenius or Hilbert-Schmidt norm. This integration allows us to obtain a dynamic circuit cost function for unitary preparation. For other mid-circuit measurement placements, we will show how to adapt the cost function in Section~\ref{sec:measure}.

Suppose we now have $s$ system qubits
and $a$ ancilla qubits, and we aim to apply a target unitary $V$ on the system qubits. As shown in Figure \ref{fig:protocol-unitary} (a), all the ancilla
qubits are initialized in $|0\rangle$ state, then we apply the $U_b$ unitary to
all qubits, followed by the simultaneous measurement of the ancilla qubits in one cycle. For each possible
measurement result, we apply a different unitary $U_i$. The protocol needs to determine $U_b$ and all $2^a$ $U_i$ branch unitaries such that the circuit corresponds to the application of the target unitary $V$ on the system qubits for any initial state $\ket{\phi}$. 
In our formulation, the unitaries $U_b$ and $U_i$ are parametrized, and we vary them until the circuit is equivalent to the target unitary $V$.

In order to build a cost function, we need to reformulate the problem. For variational unitary preparation, the Hilbert-Schmidt norm is used widely as a cost function ~\cite{cincio2021machine,davis2020towards,rakyta2022approaching,rakyta2022efficient,madden2022best}, and it is faithful and operationally meaningful as shown in \cite{khatri2019quantum,cirstoiu2020variational}. In our case, measurement is not a unitary operation, and it prevents us from using the Hilbert-Schmidt norm. For this reason, we will restate our goal by deferring the measurement as given in Figure \ref{fig:protocol-unitary} (b). For this circuit to be equivalent to the target unitary $V$, quantum state before the measurement should be $(V\ket{\phi})\otimes \ket{\alpha}$, i.e. the ancilla and the system qubits should be completely disentangled, the system qubits should be in the state $V\ket{\phi}$, and ancilla can be in an arbitrary state $\ket{\alpha}$. 
Thus, if we define
\begin{equation}
    U = \sum_{i=0}^{2^a -1} (U_i \otimes |i\rangle \langle i|)\:U_b,
\end{equation}
then it becomes clear that we would like $U$ to satisfy
\begin{align}\label{eq:unitary-prep-goal}
    U \ket{\phi} \otimes \ket{0}^{\otimes a} = (V \ket{\phi}) \otimes \ket{\alpha},
\end{align}
for an arbitrary $s-$qubit state $\ket{\phi}$. Applying the measurement as shown in Figure \ref{fig:protocol-unitary} (b), the ancilla qubits collapse to a computational basis state, and the remaining state will be $V\ket{\phi}$, which yields that we successfully implement the unitary $V$.

Defining $W\ket{0}^{\otimes a} = \ket{\alpha}$, we provide the following as a faithful cost function for Eq. \eqref{eq:unitary-prep-goal}:
\begin{align}\label{eq:Cdyn1-unitary}
    C_{\mathrm{dyn1}}(U,W) = 1 - \frac{1}{2^s} \left | \Tr(f_0 (V \otimes W)^\dagger \:U) \right |,
\end{align}
where we define $f_0 := I^{\otimes s} \otimes (|0\rangle \langle 0|)^{\otimes a}$. 
We call this cost function the dynamic circuit cost function of type 1, and the proof of its faithfulness is given in Appendix \ref{appdx:protocol-unitary}.
Note that for the edge case $a=0$, we do not have $W$ and $f_0=I^{\otimes s}$, and $C_{\mathrm{dyn1}}$ boils down to the Hilbert-Schmidt norm.

Moreover, $W$ can be optimized analytically for a given $U$. Plugging this $W$ in $C_{\mathrm{dyn1}}$, we obtain the dynamic circuit cost function of type 2 as follows (see Appendix~\ref{appdx:protocol-unitary}):
\begin{align}\label{eq:Cdyn2}
    C_{\mathrm{dyn2}}(U) = 1 - \frac{1}{4^s} \sum_i \left| \Tr(U_{i,0}\: V^\dagger) \right|^2.
\end{align}
Here we define 
\begin{equation}\label{eq:u_ij}
    U_{i,j} = \Tr_a \left( (I^{\otimes s} \otimes \bra{i}) \: U \: (I^{\otimes s} \otimes \ket{j}) \right)
\end{equation}
 as a $2^s \times 2^s$ matrix, where $\Tr_a$ is tracing over the ancilla qubits. We find that the cost function only depends on the components of $U_{i,0}$. This is because the initial state of ancilla qubits is $\ket{0^{\otimes a}}$. 
 The cost function $C_{\mathrm{dyn2}}$ can be derived directly in the open quantum systems setting by considering the target unitary and the variational dynamic circuit as two quantum channels, and using the fidelity formula given in \cite{horodecki1999general, raginsky2001fidelity}. For completeness, we provide an open quantum system based derivation in Appendix~\ref{appx:oqs}.

\begin{figure}[t]
    \centering
    \includegraphics[scale=0.45]{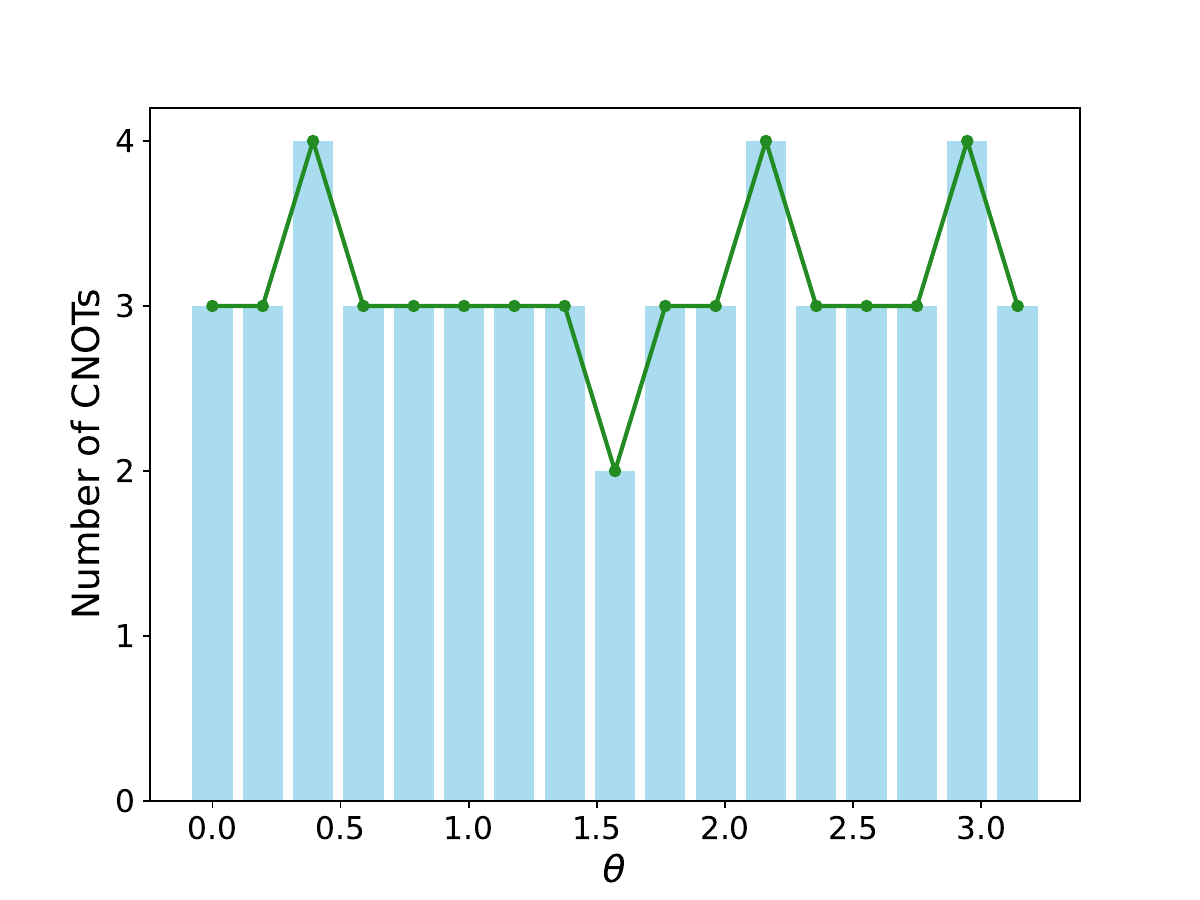}
    \caption{
    The effect of setting the measurement probabilities for the preparation of a next nearest neighbor CNOT gate by using the cost function $C_{\mathrm{dyn1}}$ given in Eq.~\eqref{eq:Cdyn1-unitary}, where we take $W = R_y(\theta)$. The CNOT count of the resulting circuit reaches a minimum value when $\theta = \pi/4$, and the measurement probabilities are equal to each other. 
    }
    \label{fig:uniform_ancilla_states}
\end{figure}

The dynamic circuit cost functions of type 1 and 2 have different features. Type 2 turns out to be more efficient than type 1 in terms of evaluation because it requires less number of arithmetic operations. It also requires less number of parameters to optimize due to the existence of the $W$ in type 1. However, the extra parameters in $W$ give type 1 the ability to control the measurement probabilities of the ancilla qubit. This can be done by fixing certain properties of $W$. For example, if we want the measurement probabilities to be uniform, we can choose  
the absolute value of the coefficients of $W\ket{0^{\otimes a}} = \ket{\alpha}$ to be equal to $1/\sqrt{2^a}$, i.e. 
%
\begin{equation}
    W \ket{0^{\otimes a}} 
    = \frac{1}{\sqrt{2^a}} \sum_{i} e^{i \phi_i} \ket{i}.
\end{equation}
Note that this can be achieved by choosing $W = D\: H^{\otimes{n}}$, where $D$ is a diagonal unitary matrix, and $H$ is the Hadamard gate, and only requires $2^a$ parameters to optimize.
In Figure \ref{fig:uniform_ancilla_states}, we illustrate this ability for a
next nearest neighbor long range CNOT gate implementation with 1 ancilla qubit,
where we set $W = R_y(\theta)$, and $W\ket{0} = \cos \theta \ket{0} + \sin \theta \ket{1}$. As it can be seen, the minimum CNOT
count is reached when $\theta = \pi/4$, where $W\ket{0} = 1/\sqrt{2}
\ket{0} + 1/\sqrt{2} \ket{1}$.
In all of our examples, we empirically observe that the optimal circuit (the circuit with the minimum depth and number of gates) is always provided by the uniform distribution of the ancillas' outcome as shown in Section~\ref{sec:eva}.

\section{State preparation}
~\label{sec:state}

Although a protocol for preparing arbitrary states using dynamic circuits has been proposed in~\cite{alam2024learning}, we present an alternative approach.
The protocol for our state preparation method using dynamic circuits is shown in Figure~\ref{fig:protocol-state}.
When we have multiple ancilla qubits for mid-circuit measurement, there are various possible measurement placements. The cost function for simultaneous mid-circuit measurements is straightforward to understand, so we will explain it first. For the cost function associated with different measurement placements, we summarize it and update our cost function accordingly in Section~\ref{sec:measure}.

Suppose we now have $s$ system qubits
and $a$ ancilla qubits, and we aim to prepare the target state
$|T\rangle$ on the system qubits. Initially, all the $s$ and $a$
qubits are in $|0\rangle$ state. We first apply the $U_b$ unitary to
all qubits, followed by the simultaneous measurement of the ancilla qubits in one cycle. For each possible
measurement result we apply a different unitary $U_i$ which takes the system qubits into the desired target state. Thus, given $|T\rangle$, the protocol needs to determine $U_b$ and all $2^a$ $U_i$ branch unitaries.
In our formulation, the unitaries $U_b$ and $U_i$ are parametrized, and we vary them until the final state of the system qubits is equal to the target state $\ket{T}$. To that effect we use a
cost function which we obtain by generalizing the fidelity between target and evolved states.

The fidelity to quantify the closeness of a pure quantum state $\ket{\phi}$ to the target state $\ket{T}$ is defined as the following
%
\begin{equation}
\label{eq:statef}
    F_T(\ket{\phi}) = |\langle T|\phi \rangle|^2.
\end{equation}
$F_T$ ranges between 0 and 1. The edge case $F_T(\ket{\phi}) = 1$ occurs if and only if $\ket{\phi} = \ket{T}$ up to a global phase, which means that to obtain the target state $\ket{T}$, we need to maximize $F_T(\ket{\phi})$. For unitary circuits that do not contain any mid-circuit measurement, $1-F_T$ is widely used as a cost function \cite{hayden2001asymptotic,kane2024nearly,gui2024spacetime}.
It is well known for the state preparation problem that this cost function is both \textit{faithful} (minimizing it guarantees a solution), and \textit{operationally meaningful} (lower values for the cost function yields less error in the state preparation)~\cite{bravo2023variational}.

In dynamic circuits, we measure ancilla qubits, and in general the final state depends on the measurement outcome. 
Therefore, we need to consider the fidelity for each final state.
Let $\ket{\phi_i}$ denote the final state that corresponds to the measurement outcome $\ket{i}$ on the ancilla qubits. 
%
Our purpose in constructing a dynamical circuit to prepare state $\ket{T}$ is to ensure $\ket{\phi_i} = \ket{T} \otimes \ket{i} = \ket{T \otimes i}$ for any $i = 0,1,\dots,2^a-1$. This can be achieved by maximizing the fidelities $F_T(\ket{\phi_i}) := |\langle T \otimes i|\phi_i \rangle|$ separately. By this intuition, we consider the following cost function:
\begin{align}\label{eq:state_prep_cost_function}
    C_T = 1 - \sum_i F_T(\ket{\phi_i}) = 1 - \sum_i |\langle T \otimes i | \phi_i \rangle|^2.
\end{align}
We prove the faithfulness and meaningfulness of this cost function in Appendix~\ref{appx:state}, and derive it from the Hilbert Schmidt norm.

For the specific case we consider in this section, the outcome states can be calculated as the following 
\begin{equation}
\begin{aligned}
    |\phi_i\rangle &= (U_i \otimes |i\rangle \langle i|) \: U_b \: |0^{\otimes (s+a)}\rangle 
\end{aligned}
\end{equation}
When we plug this in the Eq. \eqref{eq:state_prep_cost_function}, we obtain the following cost function 
\begin{equation}
    C_T(U_b, U_i) = 1- \sum_{i=0}^{2^a - 1} |\langle T \otimes i| ( U_i \otimes |i\rangle \langle i|)\: U_b\:|0^{\otimes(s+a)}\rangle|^2. 
    \label{eq:state_cost}
\end{equation}

\section{Dynamic Circuit Structures}
\label{sec:measure}
\begin{figure*}[t]
    \centering
    \includegraphics[scale=0.68]{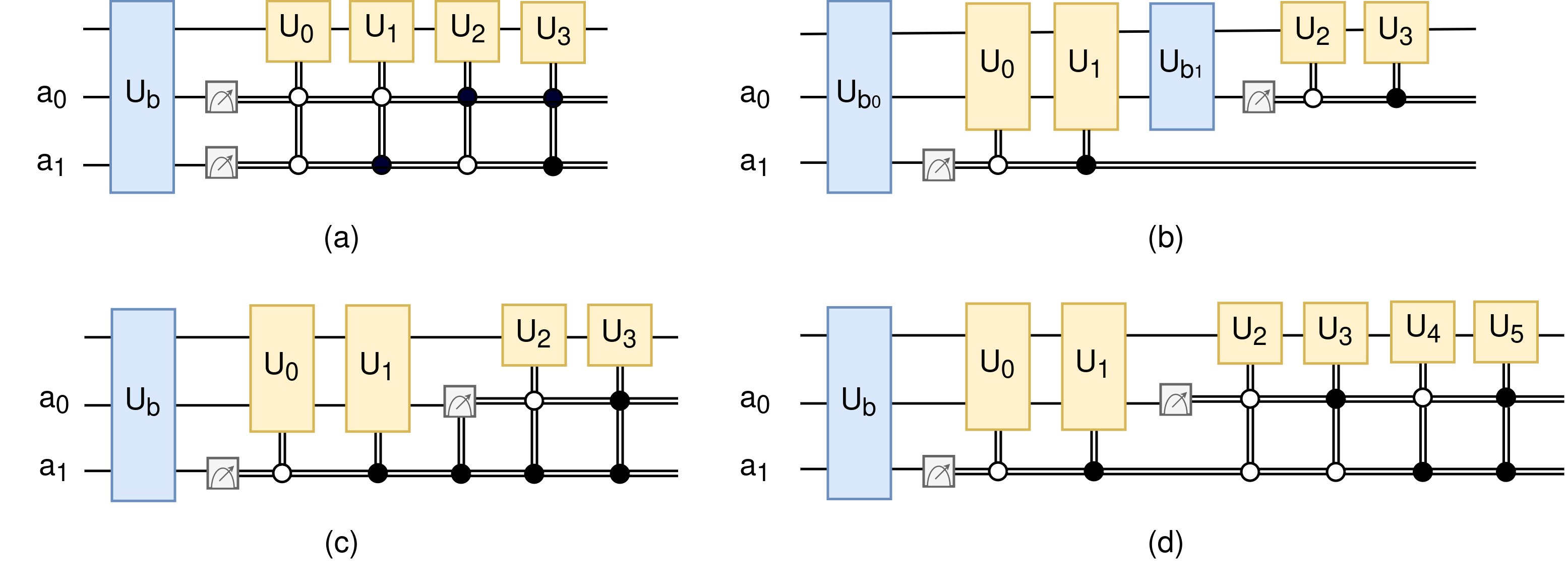}
    \caption{Circuits with two ancillas to illustrate different mid-circuit measurement scenarios. (a) Measurement of both ancillas simultaneously in one cycle. (b) Measurement of $a_1$ first, resulting in two branch circuits, followed by application of $U_{b_1}$ to the unmeasured qubits and measurement of $a_0$, creating two further branch circuits. (c) Measurement of $a_1$ first, followed by application of $U_0$ to other qubits if the measurement of ancilla qubit is $|0\rangle$. Measurement of $a_0$ only if the measurement result of $a_1$ is $|1\rangle$ state, which  creates two further branch circuits. (d) Measurement of $a_1$ first, resulting in two branch circuits, each containing a measurement of $a_0$ and unitaries applied based on the measurement outcome for $a_0$.
    }
    \label{fig:measurements}
\end{figure*}

The cost functions introduced have been derived for the
specific dynamic circuit structure in Figure~\ref{fig:protocol-state} and Figure~\ref{fig:protocol-unitary}. In
practice, when multiple measurement ancillas are used, various
dynamic circuit structures can be generated depending on measurement
placement. 
Figure~\ref{fig:measurements} illustrates possible placements for two ancilla qubits. 
Our method is generic, and the cost functions can be
easily extended for all these possible scenarios by modifying $U$ in Eq.~\eqref{eq:Cdyn1-unitary} and Eq.~\eqref{eq:Cdyn2} for unitary preparation.
For state preparation, we can update 
\begin{equation}
 \ket{\phi_i} = (I^{\otimes s}\otimes\ket{i}\bra{i}) U \ket{0}^{\otimes (s+a)}  , 
\end{equation} 
in Eq.~\eqref{eq:state_prep_cost_function}.
This full unitary $U$ can be obtained by deferring the measurement of ancillas.
For simplicity, we 
%
will derive 
$U$ for dynamic circuits with two ancillas. This method can be easily extended to scenarios involving multiple ancilla qubits.
%

\textbf{Simultaneous measurement.} As shown in Figure~\ref{fig:measurements} (a), we perform measurement on ancillas $a_0$ and $a_1$ simultaneously. This is the case which we explain in Section~\ref{sec:state} and Section~\ref{sec:unitary}. For 2 ancilla qubits, the measurement results in four branch circuits at the same time, corresponding to the four possibilities: $|00\rangle, |01\rangle, |10\rangle$, and $|11\rangle$. Depending on these outcomes, we apply $U_0, U_1, U_2$ and $U_3$ to the system qubits, respectively. The full unitary $U$ becomes:

\begin{equation}\label{eq:u_simultaneous}
    U = \sum_{i=0}^3 (U_i \otimes |i\rangle \langle i|) U_b.
\end{equation}
\textbf{Independent measurement.} As shown in Figure~\ref{fig:measurements} (b), we first measure $a_1$, resulting in two branch circuits.  We then apply $U_0$ or $U_1$ to the system qubits and the ancilla qubit $a_0$, which have not been measured yet. Following this measurement, we apply another unitary $U_{b_1}$ to all the unmeasured qubits (system qubits and $a_0$), and then measure the other ancilla qubit $a_0$. Depending on the measurement result, we apply $U_2$ or $U_3$ to only the system qubits. The full unitary $U$ becomes:

\begin{align}
\begin{split}
    U =& \left(\sum_{i=0}^1 U_{i+2} \otimes |i\rangle \langle i| \otimes I \right) \\&(U_{b_1} \otimes I)
    \left(\sum_{i=0} ^ 1 U_i \otimes |i\rangle \langle i|\right) U_{b_0}
\end{split}
\end{align}
\textbf{Asymmetric nested measurement.} Asymmetric nested measurement refers to a scenario where we first measure an ancilla qubit, resulting in two branch circuits. In one of these branches, an additional measurement is performed, leading to further branch circuits. In contrast, the other branch involves the application of a general unitary operation without any further measurements, as illustrated in Figure~\ref{fig:measurements} (c). We first measure $a_1$. Based on the measurement result, we apply $U_0$ or $U_1$ to the unmeasured qubits. Unlike independent measurement, we measure $a_0$ only if the measurement result of $a_1$ is the $\ket{1}$ state. Then we apply $U_2$ or $U_3$ to the system qubits based on whether we measure $a_0$ in $\ket{0}$ or $\ket{1}$ state, respectively. The full unitary $U$ becomes:

\begin{equation}
    \begin{aligned}
         U = & \Bigg(\sum_{i=0}^1 \left( U_{i+2} \otimes |i\rangle \langle i| \otimes I\right) \\
         & (U_1 \otimes |1\rangle \langle 1|)
         + (U_0 \otimes |0\rangle \langle 0| ) \Bigg) U_{b}
    \end{aligned}
\end{equation}

\textbf{Symmetric nested measurement.} It refers to the scenario where we first measure an ancilla qubit, resulting in branch circuits. In each of these branches, we further measure another ancilla qubit, creating additional branch circuits, as shown in Figure~\ref{fig:measurements} (d). Specifically, we start by measuring $a_1$. If the measurement result is 0, we apply $U_0$ to the unmeasured qubits, then measure $a_0$ and apply either $U_2$ or $U_3$ to the system qubits based on the outcome. Conversely, if the measurement result is 1, we apply $U_1$ to the unmeasured qubits, then measure $a_0$ again and apply either $U_4$ or $U_5$ to the system qubits based on the outcome. The full unitary $U$ becomes:

\begin{equation}
\begin{aligned}
    U =& \Bigg(\sum_{i=0} ^ 1 (U_{i+4} \otimes |i\rangle \langle i| \otimes I) \:( U_1 \otimes |1\rangle \langle 1|) \\
    &+ \sum_{i=0}^1 (U_{i+2}\otimes|i\rangle \langle i|\otimes I) \: (U_0\otimes |0\rangle \langle 0|)\Bigg) \: U_b
\end{aligned}
\end{equation}

\section{Circuit Generation Framework}

We include the novel objective functions in our AC/DC framework. For practical usage, we embed these objective functions into Berkeley Quantum Synthesis Toolkit
(BQSKit) with functionality to generate and optimize dynamic
circuits. The onus at this point is to provide a usable framework
that enables community experimentation. To demonstrate value, we
provide algorithm implementations with restricted choices of
measurement placements and circuit structures, designed to allow us to
replicate the existing published results.

We provide two new synthesis algorithms and integration into the
BQSKit large circuit optimization engine. Synthesis algorithms already
face the challenge of searching a large space of possible gate
placements, which in the dynamic circuit case is compounded by the
choices of placing the measurement operations. For this paper we
follow a simple circuit generation strategy.  Going back to
Figure~\ref{fig:protocol-state} and Figure~\ref{fig:protocol-unitary}, we need to generate the circuit for the
unitary $U_b$ and all the branch circuits associated with the
$U_i$. In the generation workflow, we use traditional
synthesis for $U_b$, then we restrict each $U_i$ to a single layer of
one-qubit gates applied to all system qubits to facilitate the evaluation of our cost function. It is straightforward to extend the $U_i$ to arbitrary unitary on the system qubits, but the searching space for synthesis can grow significantly. Overall, the approach can handle
multi-qubit, multi-ancilla circuits. 

The first algorithm,
referred to as {\it AC/DC-QSearch}, implements a bottom up circuit generation
approach based on a topology aware optimal
synthesis algorithm called QSearch~\cite{davis2020towards}.  Thus, we expect this approach to
scale to circuits with fewer than five or six qubits.  The second
approach, referred to as {\it AC/DC-Inst}, uses a top-down strategy built
based on insights gathered from generating small dynamic circuits with
{\it AC/DC-QSearch}. {\it AC/DC-Inst} scales to ten to twelve qubit circuits, which is
on par with the scalability limit of normal direct synthesis
algorithms. For large circuits, BQSKit uses a circuit partitioning~\cite{wu2020qgo}
strategy, followed by direct partition synthesis and stitching back
the optimized sub-circuits. To enable experimentation with circuit optimization at a large scale for dynamic circuits, we integrate {\it AC/DC-QSearch} and {\it AC/DC-Inst} into circuit partitioning.

{\bf AC/DC-QSearch}: We have modified the original implementation of QSearch to use
the objective functions introduced in
Sections~\ref{sec:state}~and~\ref{sec:unitary}. We also allow the
implementation to use a configurable number of ancilla qubits. We build the circuit associated with $U_b$ in
Figure~\ref{fig:protocol-state}. This prepared state is also topology aware and can take as input an user specified qubit interconnection topology. In the meanwhile we append a layer of single qubit $U_3(\phi,\theta,\lambda)$ gates on the system qubits for each branch circuit.
We keep using instantiation pass to update the parameters of $U_b$ and $U_3$ gates in the branch circuits.

{\bf AC/DC-Inst}: For the search-based synthesis algorithm applied to a
 state with a small number of qubits (no more than six),
given a linear hardware topology, we have observed a pattern in the
synthesized circuits. These circuits are composed of alternating
layers, and their depth grows approximately logarithmically, as we
will detail in Section~\ref{sec:stateres}.

Therefore, for unitary preparation in large circuits, we can use circuit partitioning to prepare each partitioned circuit using dynamic circuits in parallel, making it easily scale up to thousands of qubits, as demonstrated in~\cite{wu2020qgo}. For state preparation in large circuits, we apply a prescribed
circuit extension methodology, where we grow the circuit with fixed layers at each step.  The process is repeated until we find
the circuit that prepares the desired state. This approach allows us to scale state preparation to circuits up to 12 qubits. Also, note that since we can embed as linear topology in any other fully connected qubit topology, the approach is general.

\section{Evaluation}
~\label{sec:eva}
In this section, we first use our proposed dynamic circuit protocol to prepare various unitaries, including long-range entangled gates and multi-qubit gates.  Second, we demonstrate the application of our protocol for large quantum circuit optimization and a physical application in lattice simulation for spin systems. Moreover, we use our proposed dynamic circuit protocol to prepare various quantum states. We evaluate all the prepared dynamic circuits on both simulator and quantum hardware.
We also evaluate the fidelity trade-off between the high error rate of mid-circuit measurements and their advantages in circuit optimization, specifically in terms of circuit depth and gate count reduction. In all the evaluation, circuit depth refers to multi-qubit circuit depth. This analysis helps us understand when dynamic circuits can bring more benefits.
Some complete examples demonstrating the circuits prepared using our dynamic circuit protocol are provided in the Appendix~\ref{appdx:circuits}. Note that unless reported, all the results generated by AC/DC have errors (i.e., distance between the generated unitary/state and the target unitary/state) less than $10^{-8}$ guaranteed by the threshold set in the cost function.

\subsection{Unitary preparation}\label{sec:utryres}

In this section, we first present the results of preparing arbitrary long-range entangled gates and several multi-qubit gates. Then, we demonstrate the quantum hardware results to validate our dynamic circuit protocol.

\begin{figure*}
    \centering
    \includegraphics[width=0.99\textwidth]{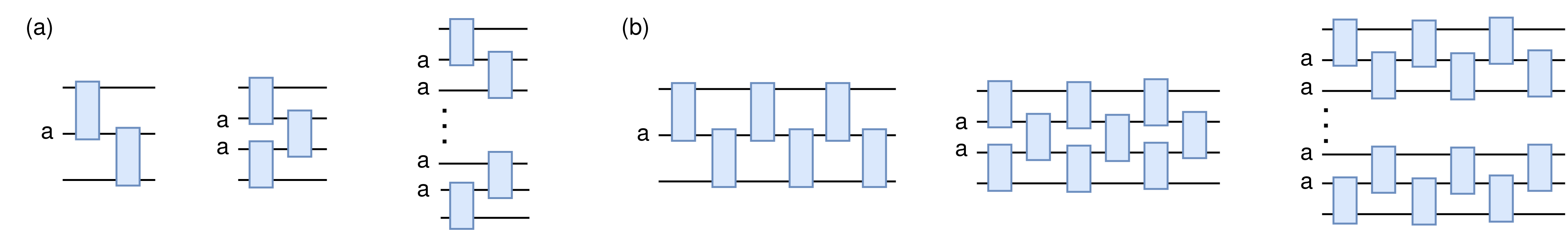}
    \caption{The circuit structures to prepare $U_b$ which occurs before mid-circuit measurement (as shown in Figure~\ref{fig:protocol-unitary}) for (a) long-range specific two-qubit gates -- CNOT, CZ, CS, Rzz, Rxx, and Ryy, and (b) long-range generic two-qubit gates, with multiple ancillas. Each blue block is explained in Eq.~\eqref{eq:block}. The mid-circuit measurement will be put on the ancilla qubits ``a'', following the branch circuits.}
    \label{fig:long-range}
\end{figure*}

\subsubsection{Long-range Two-qubit Entangling Gates}
\label{sec:long-range-gate}

First, we prepare the next nearest neighbor 2-qubit gates (with 1 ancilla in the middle): CNOT, CZ, CS, Rzz, Rxx, Ryy with arbitrary angles, and increase the number of ancillas to 2 for these long-range 2-qubit gates using our dynamic circuit protocol {\it AC/DC-QSearch}. Based on the pattern we observe, we extend it to arbitrary number of ancillas using {\it AC/DC-Inst}. The circuit structures are shown in Figure~\ref{fig:long-range} (a). We repeat the same procedure for generating long-range generic 2-qubit gates using {\it AC/DC-QSearch} and {\it AC/DC-Inst}. The circuit structures are shown in Figure~\ref{fig:long-range} (b).
One block corresponds to one CNOT and two
$U_3$ gates. Each $U_3$ gate has three independent parameters, so that one block has six parameters to optimize.
\begin{equation}
\includegraphics{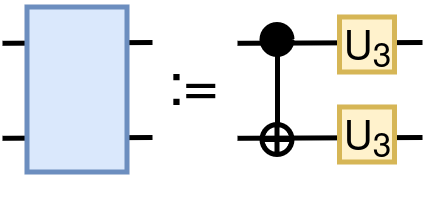}
\label{eq:block}
\end{equation}
The target topology for all the circuits are linear. 

Using this pattern, we can prepare  dynamic circuits for specific 2-qubit
long-range entangled gates (CNOT, CZ, CS, Rzz, Ryy, Rxx), where the
entangling gate acts between $q_0$ and $q_{n-1}$, using $(n-1)$ CNOTs
with depth 2. 
For a generic long-range 2-qubit gate, where the unitary can be sampled from the SU(4) group, 
a
circuit depth of 6 and $(n-1)\times 3$ CNOTs are required. 
  It has been shown in \cite{baumer2024efficient, hashim2024efficient} that the
long-range CNOT gate can be realized by gate teleportation based on
Bell states using $(n-1)$ CNOTs, which is the same number of CNOTs as
our results. It is known that an arbitrary two-qubit gate can be
realized by three CNOTs along with some single-qubit gates. This
implies that using their protocol, we can achieve a long-range
generic two-qubit gate using $(n-1)\times 3$ CNOTs and $(n-2) \times 3$ measurements. However, our
protocol can prepare the same gate using the same number of CNOTs but only $(n-2)$ measurements, reducing the number of mid-circuit measurement and
feed-forward loop by a factor of three (an example circuit of preparing an arbitrary two-qubit gate using a three-qubit circuit with one ancilla is shown in Figure~\ref{fig:dc-circuits} (d)). Since measurement and
feed-forward loops are the operations that take the longest time and
are the most error-prone in current quantum hardware, this improvement
can directly enhance the circuit fidelity significantly.

We compare our dynamic circuits results, labeled as AC/DC, with the
following: (1) We construct a $n$-qubit circuit with long-range generic two-qubit gate on $(q_0, q_{n-1})$ and use Qiskit with optimization level 3 to compile it to linear topology. (2) We use Quantum Shannon Decomposition (QSD) to decompose a $2^n \times 2^n$ unitary matrix for long-range generic two-qubit gate. We use method as
implemented in Qiskit's synthesis module and transpiled the resulting
circuits to linear connectivity with optimization level 3. 

\begin{figure}
    \centering
    \includegraphics[scale =0.28]{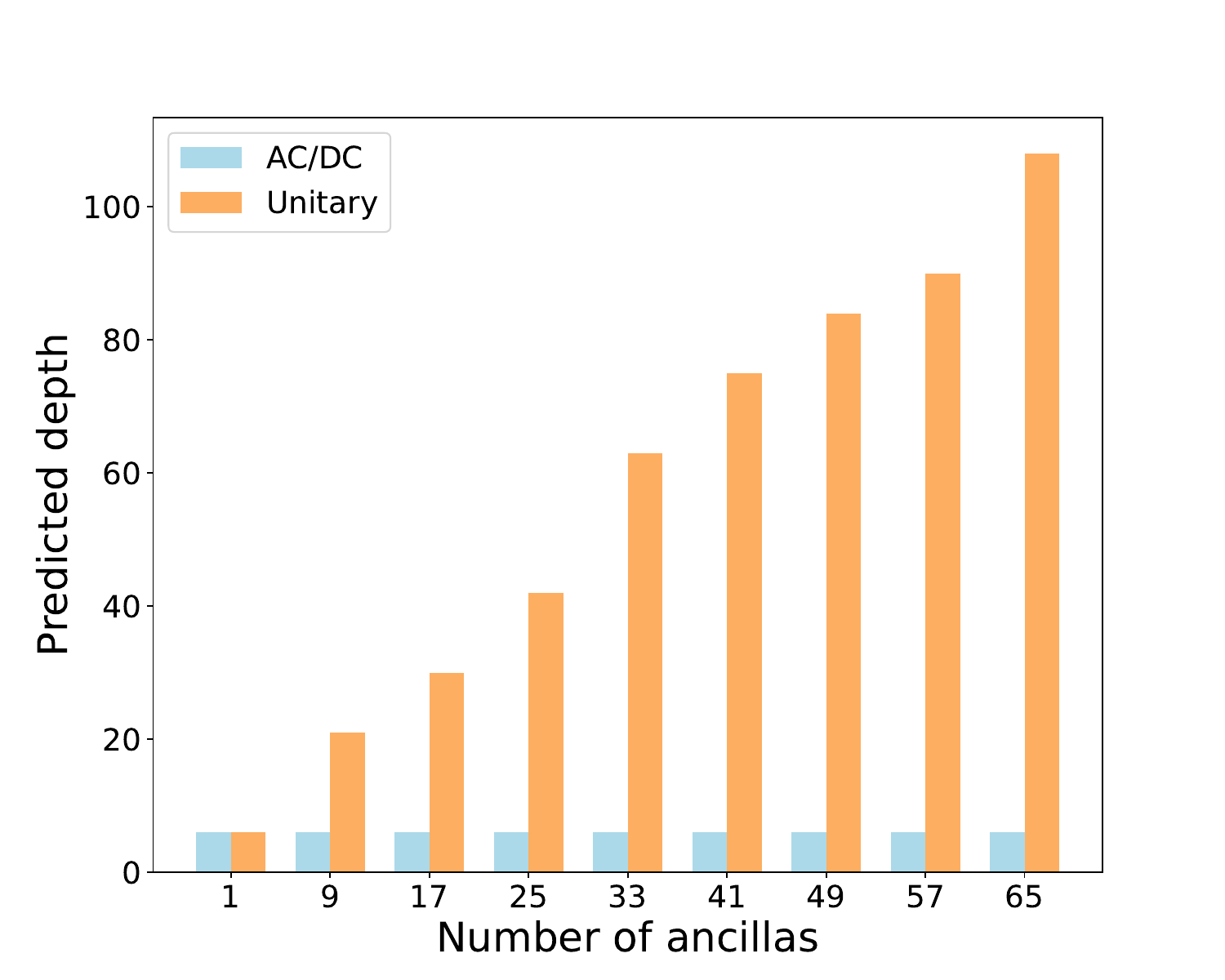}
    \caption{The predicted circuit depth of long-range generic two-qubit gates prepared by dynamic circuits (AC/DC) and unitary circuits. The total number of qubits is the number of ancillas plus 2 data qubits, with the long-range two-qubit gate applied to the first and last qubits (data qubits) and the ancilla qubits in between.}
    \label{fig:long-range-depth}
\end{figure}

When preparing a long-range generic two-qubit unitary, Qiskit and AC/DC give the same number of CNOTs. However, QSD requires no more than $(23/48) \times 4^n -(3/2) \times 2^n + 4/3$ CNOTs according to~\cite{shende2005synthesis}, which is significantly larger than the other two methods. For example, implementing a long-range generic two-qubit gate in a 10-qubit circuit requires approximately 500,000 CNOT gates using QSD, whereas it only takes 27 CNOT gates with Qiskit and AC/DC. In Figure ~\ref{fig:long-range-depth}, we show the circuit depth comparing AC/DC and unitary circuits prepared by Qiskit. Note that the AC/DC results for large quantum circuits are based on predictions derived from observed patterns. AC/DC has constant depth of 6, while the depth of unitary circuits is increased with the number of ancillas.

\subsubsection{Multi-qubit Gates}

Another application of dynamic circuits is to introduce ancilla qubits for unitary preparation, allowing the use of more qubits to prepare a smaller unitary with the potential to reduce depth to achieve circuit optimization. For demonstration, we select several three-qubit gates, including Toffoli, Fanout, and Fredkin gates, and introduce one ancilla qubit. Using four-qubit dynamic circuits with one ancilla, we prepare the 3-qubit unitaries on a linear topology. 

We compare the dynamic circuits with preparing the 3-qubit unitaries directly using BQSKit and QSD without any ancilla qubits. The results are summarized in Table~\ref{tab:multi-qgate}. Note that in the four-qubit dynamic circuits, any one of the four qubits can be designated as an ancilla qubit, resulting in four possible configurations. Each configuration introduces a different number of CNOT gates. In our results, we present the configuration that yields the shortest circuit depth. By
preparing the smaller quantum gate in a larger quantum circuit, AC/DC
approach achieves the shortest circuit depth among the methods while
slightly increase the number of CNOT gates compared with BQSKit. In contrast, QSD
consistently produces the poorest results, exhibiting both increased
gate counts and longer circuit depths. These small benchmarks demonstrate the potential and trade-offs of increasing circuit width to reduce circuit depth.

\begin {table}[t]
\begin{center}
\caption{Preparation of three-qubit gates using AC/DC, BQSKit and QSD on linear topology. Note that, AC/DC involves one ancilla qubit so the circuit size is four. For the rest of the results using BQSKit and QSD, the circuit size is three. AC/DC achieves the shortest circuit depth.}
\label{tab:multi-qgate}
\footnotesize
\begin{tabular}{|c|c|c|c|c|c|c|}
\hline  
\multirow{2}{*}{Gate} & \multicolumn{3}{c|}{CNOT} & \multicolumn{3}{c|}{Depth} \\
\cline{2-7}
 & AC/DC & BQSKit & QSD & AC/DC & BQSKit & QSD \\
\hline
Toffoli & 9 & 7 & 30 & 6 & 7 & 30 \\
\hline
Fanout & 4 & 3 & 28 & 3 & 3 & 28\\
\hline
Fredkin & 11 & 9 & 29 & 8 & 9 & 29 \\
\hline
\end{tabular}
\end{center}
\label{tab:multi-qubit}

\end{table}

\begin{figure*}[htbp!]
    \centering
    \includegraphics[width=\textwidth]{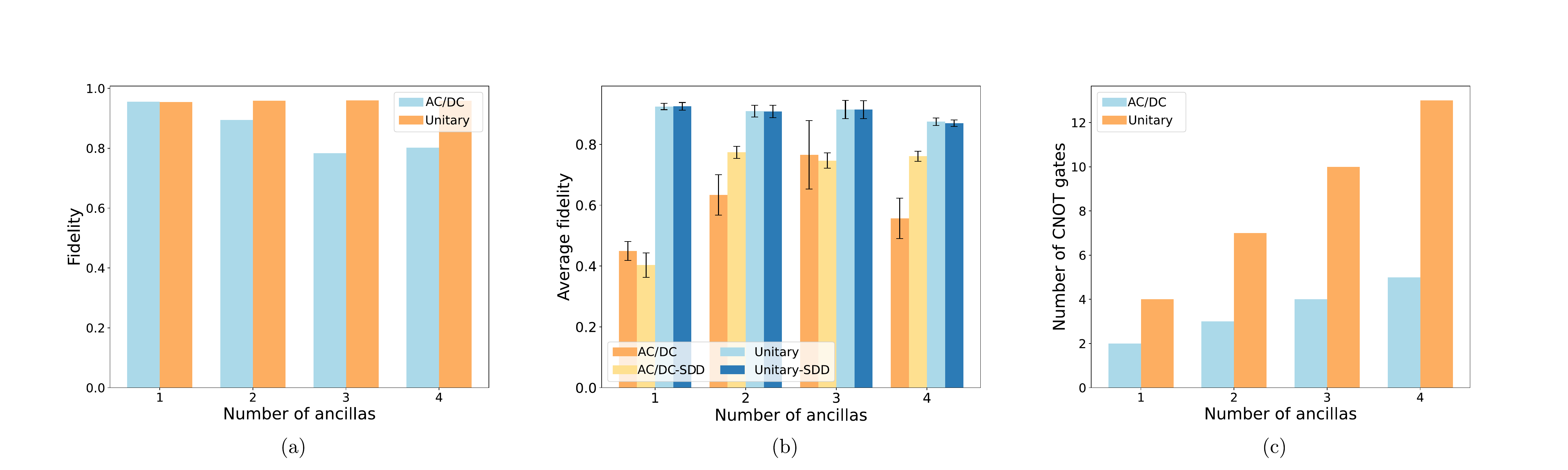}
    \caption{     
    The fidelity results for executing unitary and dynamic circuits (AC/DC) to prepare long-range CNOT gate on linear chain on (a) noise simulator (b) IBM quantum hardware. 
    (c) The number of CNOTs required for circuit preparation. 
    The total number of qubits is the
number of ancillas plus 2 data qubits, with the
long-range CNOT applied to the first and last
qubits (data qubits) and the ancilla qubits in between. SDD represents staggered dynamical decoupling.}

    \label{fig:cx-fid}
\end{figure*}

\subsubsection{Validation of Generated Circuits}
\label{sec:unitary-fid}

For validation, we use dynamic circuits and unitary circuits (Qiskit with optimization level 3) to prepare the circuits for long-range CNOT gates with ancilla qubits ranging
from one to five, as well as a Toffoli gate with one ancilla. We run the circuits on Qiskit noise simulator {\it Fake127QPulseV1} and IBM quantum hardware {\it ibm\_nazca}. For the hardware execution, we map the circuits to three linear partitions of the quantum hardware
to remove the bias on certain qubit measurement error.
To reduce the idling errors, we apply staggered dynamical decoupling (SDD) to the circuits~\cite{niu2024multi}.

We employ the truth table tomography to validate the unitary preparation~\cite{hashim2024practical},
where the fidelity $F$ is defined as:
\begin{equation}
    F = \frac{1}{2^s} \Tr(S_{\text{exp}}^T S_{\text{ideal}}),
\end{equation}
where $s$ represents the Hilbert space dimension for a quantum system with $s$ system qubits. $S_{\text{ideal}}$ is the target unitary matrix and $S_{\text{exp}}$ is a matrix constructed by the output results corresponding to initial states in different computational basis states. For example, if the target is a CNOT gate,
we prepare the circuits for long-range CNOT in
computational basis states:  $|00\rangle, |01\rangle,
|10\rangle,$ and $|11\rangle$. Similarly, for a Toffoli gate, the
initial states are prepared in all eight basis states, from
$|000\rangle$ to $|111\rangle$. 
$S_{\text{ideal}}$ is the CNOT or Toffoli matrix, and we compare it with the
experimental results $S_{\text{exp}}$.

The fidelity results and the number of CNOT gates to prepare the long-range CNOT gates are shown in Figure~\ref{fig:cx-fid}. Even though dynamic circuits reduce the number of CNOTs by 59\% compared to unitary circuits, the noisy simulation results show that unitary circuits achieve higher fidelity due to the high errors of mid-circuit measurements (MCM) in AC/DC. For a small number of ancillas, the fidelity of AC/DC decreases as the number of ancillas increases, while the fidelity of unitary circuits remains relatively stable. However, for the hardware results, the fidelity of unitary circuits starts to drop when the total number of qubits reaches six. Given the high MCM error, it is surprising that the configuration with the fewest mid-circuit measurements and feed-forward loops yields the worst fidelity results. Initially, fidelity increases as we increase the number of ancilla qubits, but then it drops. SDD significantly improves the fidelity of dynamic circuits, as these circuits have longer idle times due to the mid-circuit measurement and feed-forward loop. We observe the same behavior for the Toffoli gate. After applying SDD, even though the dynamic circuit has a shorter circuit depth but the same number of CNOTs as the unitary circuits, the fidelity is only 0.31, while the unitary circuit achieves 0.84.

Based on the high errors of MCM and the hardware results, an interesting future direction is to incorporate MCM error into the design of dynamic circuits. However, our primary contribution focuses on cost function design. While the extension to an error-aware compiler for MCM is a promising avenue for future work, it falls outside the scope of this paper.

\subsection{Large Circuit Optimization}

\begin{table*}[t]
\begin{center}
\caption{Circuit partition results using AC/DC, BQSKit and a mixed method. Both mixed and AC/DC results have the shortest circuit depth but mixed method requires less number of qubits and CNOTs compared with AC/DC. Note that the number of qubits for AC/DC can be significantly reduced if we enable qubit reuse technique.}
\label{tab:partition}
\footnotesize
\begin{tabular}{|c|c|c|c|c|c|c|c|c|c|c|c|c|}
\hline  
\multirow{2}{*}{Circuits} & \multicolumn{4}{c|}{$n$} & \multicolumn{4}{c|}{CNOT} & \multicolumn{4}{c|}{Depth}  \\
\cline{2-13}
  & Original  & AC/DC &  BQSKit & Mixed & Original & AC/DC & BQSKit & Mixed & Original & AC/DC & BQSKit & Mixed\\
\hline
Grover & 6 & 38 & 6 & 33 & 174 & 171 & 165 & 161 & 172 & 101 & 165 & 101\\
\hline
tfim-2-10 & 10 & 20 & 10 & 17 & 36 & 43 & 35 & 43 & 22 & 28 & 35 & 28\\
\hline
tfim-2-12 & 12 & 22 & 12 & 21 & 44 & 46 & 37 & 46 & 26 & 29 & 37 & 29\\
\hline 
QAOA-random & 10 & 25 & 10 & 20 & 54 &  66 & 53 & 63 & 26 & 43 & 53 & 43\\
\hline
QAOA-3-regular & 10 & 19 & 10 & 16 & 30 & 38 & 32 & 38 & 18 & 25 & 32 & 25\\
\hline
\end{tabular}%
\end{center}
\end{table*}

We demonstrate the potential of employing dynamic circuits in large-scale circuit optimization. For this experiment, we configure the BQSKit infrastructure to partition circuits into three-qubit blocks. We explore this partitioning method on several quantum algorithms, including Grover's Algorithm, the Transverse Field Ising Model (TFIM), and the Quantum Approximate Optimization Algorithm (QAOA). Note that TFIM includes two Trotter steps, and QAOA circuits are constructed based on both a random graph and a 3-regular graph.

For each three-qubit block, we first use BQSKit (Qsearch) to synthesize it. Then, we replace each block with a four-qubit dynamic circuit with one ancilla and use our dynamic circuit protocol to synthesize it. Each block is synthesized in an all-to-all connectivity. After obtaining all the synthesized circuits, we explore a mixed method, choosing the synthesized block between dynamic circuits and BQSKit with the best combination of depth and CNOT count. Specifically, we prioritize the block with the shorter circuit depth. If the circuit depths are the same, we choose the one with the fewer number of CNOTs. The results are shown in Table~\ref{tab:partition}.

Dynamic circuits reduce the circuit depth on average by 30\% compared to
BQSKit synthesis. On the other hand, the number of CNOTs slightly
increased for some benchmarks, on average by 13\%. We attribute this increase to lack of integration of
the mapping BQSKit pass and believe it can be improved.  As
mentioned, the circuit width increases with the number of three-qubit
partitions, which can be mitigated using the mixed method. 
It achieves the same circuit depth as AC/DC while reducing the CNOT count and number of ancillas. For example, the mixed method produces the smallest number of CNOTs and the
shortest circuit depth for the Grover circuit among all the methods.


The current implementation of applying dynamic circuit for large circuit optimization is naive and serves as a proof of
concept.  We introduce one ancilla per block, resulting in an increase
in the number of qubits proportional to the circuit depth (total
number of blocks), rather than proportional to original width (the
number of blocks running in parallel). The number of ancillas will be reduced significantly if we enable ancilla qubit reuse technique. We also do not enable the
sophisticated~\cite{liu2023tackling} mapping optimizations within
BQSKit. Nevertheless, we believe the results are already promising and
showcase the potential of this optimization technique. Some of these trends can be already easily remediated by employing a
hybrid synthesis approach where each block is re-synthesized using the
general and the dynamic circuit approach. These results demonstrate how our dynamic circuit unitary preparation
method can be scaled up for large circuit optimization, with already promising potential. We anticipate
that a more fine-tuned method of combination and ancilla reuse can be very competitive.

\subsection{Simulation of Spin Systems}

\begin{figure*}
    \centering
    \includegraphics[width=\linewidth]{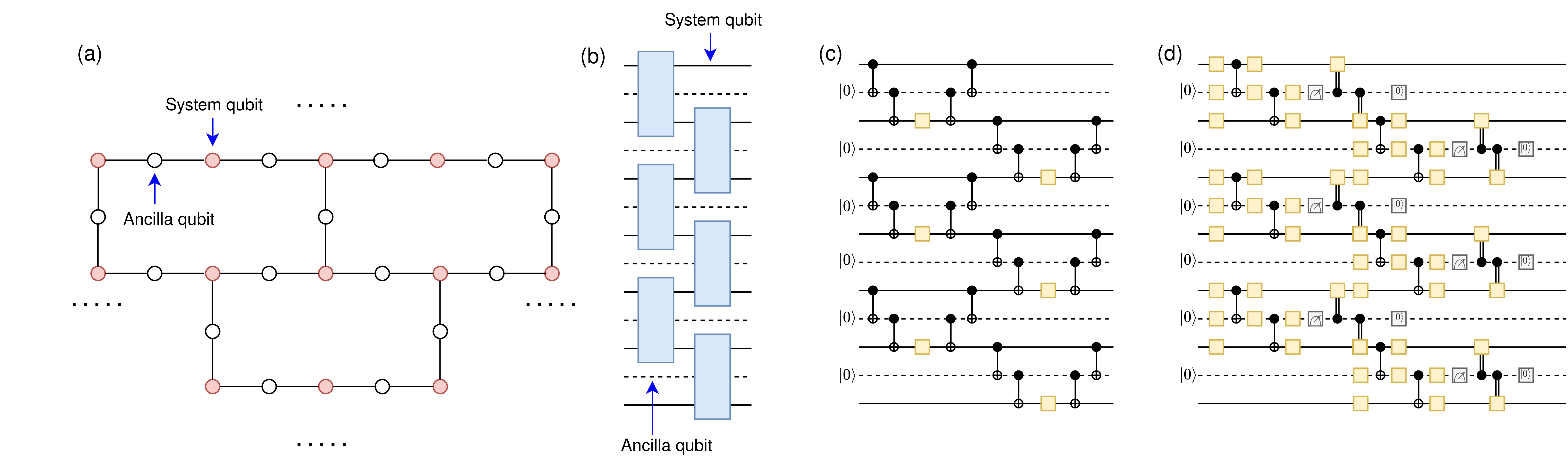}
    \caption{
    On panel \textbf{(a)}, we illustrate the qubit mapping for the TFIM Hamiltonian on a hexagonal lattice to a heavy-hexagonal architecture. This mapping turns out to be the optimal mapping for large-scale hexagonal TFIM. Panel \textbf{(b)} illustrates Rzz component of one Trotter step of the time evolution circuit of the TFIM Hamiltonian. Due to the mapping, Rzz gates must be applied between nearest neighbor system qubits, which are represented by blue rectangles. Since the ancilla qubits are not used, they can always be set to $\ket{0}$ state.  Note that $R_x$ gates are ignored as our focus is on qubit mapping. In panel \textbf{(c)}, we take advantage of this fact, and implement this Trotter step by only using unitary gates with BQSKit-0 (using cost function in Eq.~\eqref{eq:bqskit-0}), which results in a circuit depth of 7 CNOTs per Trotter step. Note that the ancillas remain in $\ket{0}$ state before and after the application of the Trotter step. We perform gate deletion to remove some redundant single-qubit gates. Finally in panel \textbf{(d)}, we implement the same circuit using dynamical circuits via optimizing our cost function Eq.~\eqref{eq:Cdyn2} (DC-Rzz), which lead to a circuit depth of only 4 CNOTs and 2 cycles of measurements. At the end, we reset the ancilla qubits to ensure they remain in $\ket{0}$ state.
    }
    \label{fig:tfim-cir}
\end{figure*}

In this section, we demonstrate the application of dynamic circuits
for unitary preparation in simulating spin systems on selected
lattice structures. For example, if we simulate time evolution of the
TFIM on a hexagonal lattice, the Hamiltonian is given by:
\begin{equation}
    H = J \sum_{(i, j) \in E} Z_iZ_j + B\sum_{i \in V} X_i
    \label{eq:tfim-ham}
\end{equation}
where $J$ is the coupling strength, $B$ is the magnetic field, and $ G
= (E, V)$ represents the target hexagonal lattice with $E$ as the set
of edges and $V$ as the set of vertices. In this context, $Z_iZ_j$ are
the interactions between nearest neighbor.

When simulating this
Hamiltonian on a quantum computer with limited qubit connectivity,
such as the heavy-hex architecture used by IBM quantum hardware, a
natural and scalable approach is to map the time-evolution circuit qubits to next
nearest-neighbor physical qubits, as illustrated by the red dot in
Figure~\ref{fig:tfim-cir} (a). This way, the connectivity between next
nearest-neighbor qubits can be managed using a long-range Rzz gate
with an ancilla qubit in between. Then we can prepare the long-range Rzz gate using dynamic circuit. 

Alternatively, we can also prepare this long-range Rzz gate using a unitary circuit. To do so, we introduce a unitary synthesis method, BQSKit-0, which is a variant of BQSKit that generates unitaries
while setting the initial state of the middle (ancilla) qubits to $\ket{0}$. We compare dynamic circuits with BQSKit-0 because dynamic circuits also require the initial states of the ancilla qubits to be $\ket{0}$. By adding this constraint, the number of CNOTs required to prepare a long-range Rzz gate is smaller than directly using BQSKit or Qiskit for long-range Rzz preparation. In this
case, the cost function for instantiation is changed from the
Hilbert-Schmidt distance to:
\begin{equation}
    C = 1 - \frac{|\Tr(f_0 U V^{\dagger})|}{\Tr(f_0)},
    \label{eq:bqskit-0}
\end{equation}
where $V$ is the target unitary matrix, $U$ is the prepared unitary
matrix, and $f_0$ represents the operation of setting $|0\rangle
\langle0|$ on the middle qubits and identity on the other qubits to
ensure that the middle qubits always start in the state
$|0\rangle$, which is similar as Eq.~\eqref{eq:Cdyn1-unitary}. This lattice structure aware mapping method is straightforward and can
be scaled to large circuits with an arbitrary number of qubits. For one Trotter step, we need $(4\times E)$ CNOT gates when using BQSKit-0, and $(2\times E)$ CNOT
gates if we use our dynamic circuit approach to optimize the circuits,
where $E$ is the number of edges in the graph. 

\begin{figure}
    \centering
    \includegraphics[scale=0.25]{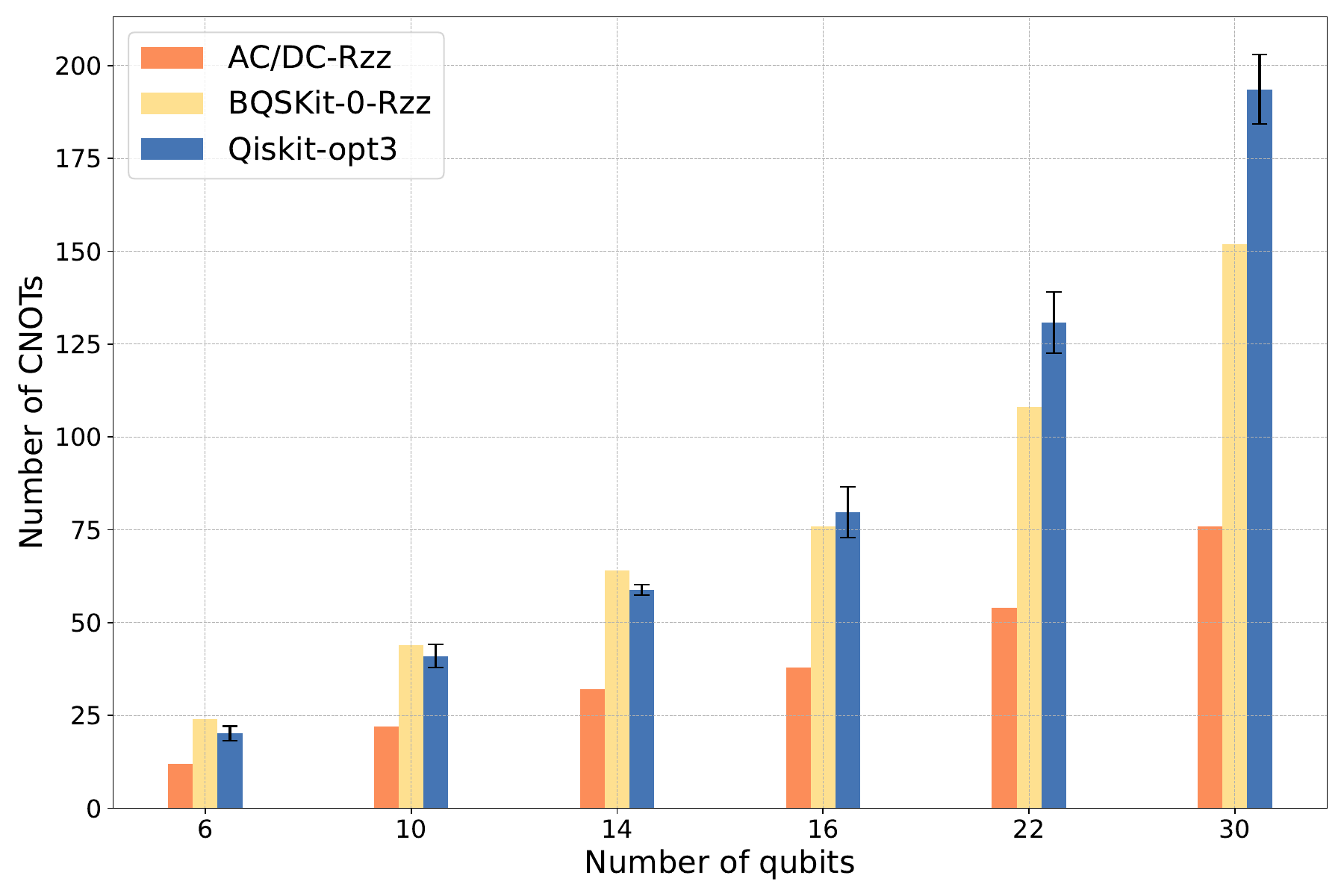}
    \caption{Number of CNOTs when mapping the TFIM circuits based on hexagonal lattice to the heavy-hexagonal architecture using different methods.}
    \label{fig:lattice-cnot}
\end{figure}

Instead of mapping the circuits to the heavy-hex architecture
according to the lattice structure, we can also use a general compiler
for the mapping, regardless of the lattice structure. We construct our
circuits based on the hexagonal lattice with different numbers of
nodes, corresponding to different numbers of qubits, and map the
circuits to the heavy-hex architecture using Qiskit with
optimization level 3.

Overall, we use the lattice structure-aware method by synthesizing the
long-range Rzz gate using BQSKit-0 (labeled as BQSKit-0-Rzz), dynamic circuits (labeled as DC-Rzz), and
compare the results with circuits compiled directly using Qiskit (labeled as Qiskit-opt3). The Qiskit
results are repeated five times due to the variance. The results are shown in
Figure~\ref{fig:lattice-cnot}.

The dynamic circuit always results in the smallest number of CNOTs, reducing the number of CNOTs by 50\% and 55\% compared with the lattice structure-aware mapping (BQSKit-0-Rzz) and the general compiler (Qiskit), respectively. Moreover, the time to find the mapping for a quantum circuit using a general compiler usually increases with the circuit size, whereas the lattice structure-aware mapping and dynamic circuit methods are scalable and independent of the system size. This example shows that general compilers can only generate good mapping strategies for small-sized circuits. As the number of qubits increases, the general compiler takes a long time to find a mapping, and it is not as effective as directly using the scalable and fast lattice structure-aware mapping method. 
Additionally, the circuit depths produced by general compilers increase with the number of qubits, whereas the AC/DC consistently yields a circuit depth of 4 and the BQSKit-0 yields a constant circuit depth of 7, whose circuit structures are shown in Figure~\ref{fig:tfim-cir} (c)-(d). The reduction in CNOT depth comes at the cost of an increased measurement depth. Even though the measurement operation is longer than the CNOT gate, as we improve the hardware implementation of MCMs, the dynamic circuit version for lattice simulation will eventually bring benefits. We can also explore the combination of BQSKit-0 and AC/DC to balance the increase of MCM number and reduction of circuit depth and gate number.

While preparing our manuscript, we learned that a similar experiment using dynamic circuits for lattice simulation was demonstrated by the IBM team~\cite{qiskit2024}. Their approach is analytic, whereas we employ our AC/DC numerical method to identify a similar structure. The advantage of AC/DC lies in its numerical flexibility; we can utilize lattice embeddings to generate dynamic circuits for arbitrary lattice structures and various condensed matter physics models.

\subsection{State preparation}\label{sec:stateres}

In this section, we present the results of preparing GHZ and W states
and  compare them with the
state-of-the-art techniques. Additionally, we demonstrate the hardware
results and explain the protocol used to validate the states prepared
on the quantum hardware.

\subsubsection{GHZ state}
\label{sec:ghz}
\begin{figure*}[t]
    \centering
    \includegraphics[scale=0.8]{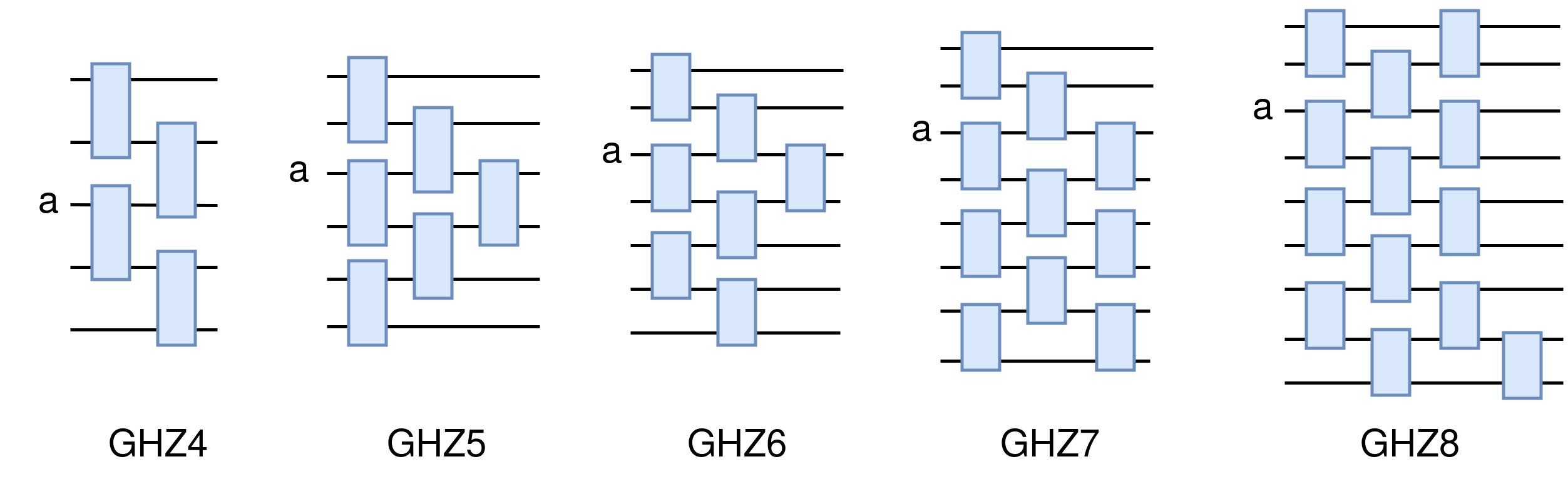}
    \caption{ The circuit structures to prepare $U_b$ (as shown in Figure~\ref{fig:protocol-state}) for GHZ4-GHZ8 states, which occurs before the mid-circuit measurement. Each blue block is explained in Eq.~\eqref{eq:block}. The mid-circuit measurement will be put on the ancilla qubit ``a'', following the branch circuits. }

    \label{fig:ghz_circuits}
\end{figure*}

A $n$-qubit GHZ state is defined as:
\begin{equation}
    |\mathrm{GHZ}_n\rangle = \frac{|0\rangle^{\otimes n} + |1\rangle^{\otimes n}}{\sqrt{2}}
\end{equation}
We first prepare 4-8 qubit GHZ states using dynamic circuits with one
ancilla qubit, resulting in 5-9 qubit circuits. We use {\it AC/DC-QSearch} for circuits smaller than 6 qubits
and an {\it AC/DC-Inst} for larger
circuits. The target topology is linear. The circuits to prepare $U_b$ before mid-circuit measurement and branch circuits using our dynamic circuit protocol are shown in Figure
\ref{fig:ghz_circuits}. 

The unitaries for branch circuits are one layer of $U_3$ gates so we only show the structure of the main circuit $U_b$. Note that $U_b$ starts with an initial layer containing one $U_3$ gate on each qubit, it is ignored in the circuit structure in the figures since we focus on multi-qubit circuit depth.
We keep the
same convention for figures representing other circuit structures. 
The structures of dynamic circuits for GHZ preparation follow a consistent
pattern, which makes it feasible to use {\it AC/DC-Inst} directly.
The circuit depth
and number of CNOTs vary when we choose different ancilla qubits, as
our target is linear connectivity. We observed that setting the
ancilla qubit to $q_2$ consistently yields the best results, so we
only present the corresponding circuits and results here.

The results are compared with: (1) the default state preparation method used in Qiskit~\cite{iten2016quantum}. We use the Qiskit implementation to generate the circuits with gate sets composed of CNOT and $U_3$ with optimization level
3; (2) A practically efficient method~\cite{github_repo} for constructing GHZ state
circuits based on the target hardware topology, and labeled as
E-GHZ; (3) BQSKit with
optimization level 3.

The results are shown in Table \ref{tab:ghz}. We present the number of
CNOTs and multi-qubit circuit depth. In terms of the time consumption, the Qiskit default method and E-GHZ operate at millisecond scale, as these are analytic methods. While BQSKit and AC/DC function on the hours scale, taking longer due to their numerical techniques. We set a threshold of ten hours, marking results as N/A when the processing time exceeds this limit. However, it is important to note that time is not the measure for success in our AC/DC framework. The primary emphasis lies on the ability to utilize this cost function for state preparation using dynamic circuits.
Regarding gate count and circuit depth, the Qiskit default method yields the poorest results among all
the methods. Both E-GHZ and BQSKit perform well in terms of the number
of CNOTs and circuit depth. Among all these methods, dynamic circuit state preparation
achieves the shortest circuit depth, with the slowest growth in
circuit depth, albeit at the cost of more CNOTs.

We also use a 7-qubit circuit including 2 ancillas to prepare the 5-qubit GHZ
state. The two measurements are performed simultaneously as shown in Figure~\ref{fig:measurements} (a). The synthesized
circuit has 6 CNOTs with depth 2, which has the same amount of CNOTs
compared with the dynamic circuit with 1 ancilla, but shorter depth.

\begin{table}[t]
\begin{center}
\caption{Number of CNOTs and circuit depth when preparing circuits for GHZ using Qiskit default method, E-GHZ, BQSKit, and AC/DC with one ancilla qubit. AC/DC achieves the shortest circuit depth with an increase in CNOTs. It can further reduce the circuit depth if more ancillas are introduced.}
\label{tab:ghz}
\footnotesize
\begin{tabular}{|c|c|c|c|c|c|c|c|c|}
\hline  
GHZ & \multicolumn{4}{c|}{CNOT} & \multicolumn{4}{c|}{Depth} 
\\
\hline
$n$ & \makecell{Qiskit \\ default} & E-GHZ & BQSKit & \makecell{AC/DC \\ (1 anc)} & \makecell{Qiskit \\ default} & E-GHZ & BQSKit & \makecell{AC/DC \\ (1 anc)}\\
\hline
4 & 25 & 3 & 3 & 4 & 24 & 3 & 2 & 2\\
\hline
5 & 62 &  4 & 5 & 6 & 57 & 3 & 3 & 3\\ 
\hline
6 & 140 & 5 & 6 & 7 & 115 & 4 & 4 & 3\\
\hline
7 & 305 & 6 & N/A & 10 & 216 & 4 & N/A & 3\\
\hline
8 & 656 & 7 & N/A & 13 & 445 & 5 & N/A & 4\\
\hline 
\end{tabular}

\end{center}
\end{table}

\subsubsection{W state and Dicke state}
\label{sec:w}
A $n$-qubit W state is defined as:
\begin{equation}
    |\mathrm{W}_n\rangle = \frac{1}{\sqrt{n}}\big(|100...0\rangle + |010..0\rangle + ... + |00...01\rangle \big)
\end{equation}
For example, if $n$ is 3, we get $|\mathrm{W}_3\rangle = \frac{1}{\sqrt{3}} (|100\rangle + |010\rangle + |001\rangle)$. The W state can be regarded as a special case of the the Dicke state with $k=1$, where a general Dicke state is defined as:
\begin{equation}
 |D_k^{n}\rangle = \binom{n}{k}^{-\frac{1}{2}}\sum_{\substack{x \in \{0, 1\}^n \\ \mathrm{HW}(x)=k}}|x\rangle ,
\end{equation}
where HW represents the Hamming weight, indicating the number of 1s in the bitstring $x$. 

\begin{figure}
    \centering
    \includegraphics[scale =0.9]{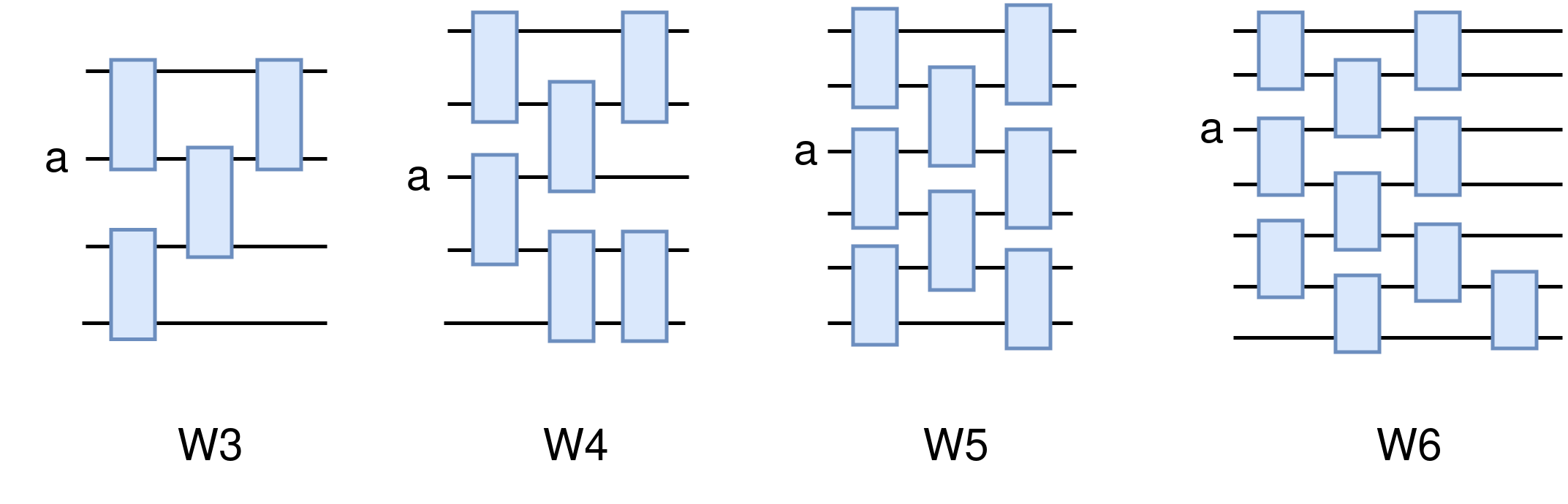}
    \caption{ The circuit structures to prepare $U_b$ (as shown in Figure~\ref{fig:protocol-state}) for W3-W6 states, which occurs before the mid-circuit measurement. Each blue block is explained in Eq.~\eqref{eq:block}. The mid-circuit measurement will be put on the ancilla qubit ``a'', following the branch circuits.}
    \label{fig:w_circuits}
\end{figure}

We use {\it DC-QSearch} and {\it DC-Inst} to prepare 3-6 qubit W states with one ancilla for linear topology, resulting in 4-7 qubit circuits. The circuit structures are shown in Figure~\ref{fig:w_circuits}. Additionally, we prepare the Dicke state $|D_2^3\rangle = \frac{1}{\sqrt{3}}(|011\rangle + |101\rangle + |110\rangle)$ and $|D_2^4 \rangle = \frac{1}{\sqrt{6}} (|0011\rangle + |0101\rangle + | 0110\rangle + | 1001\rangle + | 1010\rangle + |1100\rangle)$.

Similarly to the GHZ experiment, we compare the results with the Qiskit default state preparation method with optimization level 3 and BQSKit with optimization level 3. The results are shown in Table~\ref{tab:w}. The Qiskit default state preparation method consistently yields the worst results, while the dynamic circuit approach provides the shortest circuit depth and the slowest growth in circuit depth.



\begin {table}[t]
\begin{center}

\caption{Number of CNOTs and circuit depth when preparing circuits for W and Dicke states using Qiskit default method, BQSKit and AC/DC with one ancilla. AC/DC achieves the shortest circuit depth. It can further reduce the circuit depth if more ancillas are introduced.}

\label{tab:w}
\footnotesize
\begin{tabular}{|c|c|c|c|c|c|c|c|}
\hline  
\multirow{2}{*}{States} & \multirow{2}{*}{$n$} & \multicolumn{3}{c|}{CNOT} & \multicolumn{3}{c|}{Depth} 
\\
\cline{3-8}
& & \makecell{Qiskit \\ default} & BQSKit & \makecell{AC/DC\\ (1 anc)} &  
\makecell{Qiskit \\ default} & BQSKit & 
\makecell{AC/DC \\ (1 anc)}  \\
\hline
$\mathrm{W}_3$ & 3 & 8 & 3 & 4 & 8 & 3 & 3\\
\hline
$\mathrm{W}_4$ & 4 & 24 & 5 & 6 & 23 & 4 & 3\\
\hline
$\mathrm{W}_5$ & 5 & 65 & 8 & 8 & 52 & 5 & 3\\
\hline
$\mathrm{W}_6$ & 6 & 144 & 9 & 10 & 113 & 7 & 4\\
\hline
$D_2 ^ 3$ & 3 & 7& 3 & 4 & 7 & 3 & 3\\
\hline
$D_2 ^ 4$ & 4 & 23& 11 & 9 & 22 & 7 & 5\\
\hline %
\end{tabular}

\end{center}
\end{table}

\subsubsection{State Preparation Validation}
\label{sec:state-fid}
\begin{figure*}[htbp!]
    \centering
    \begin{subfigure}{0.45\textwidth}
    \centering
        \includegraphics[scale=0.2]{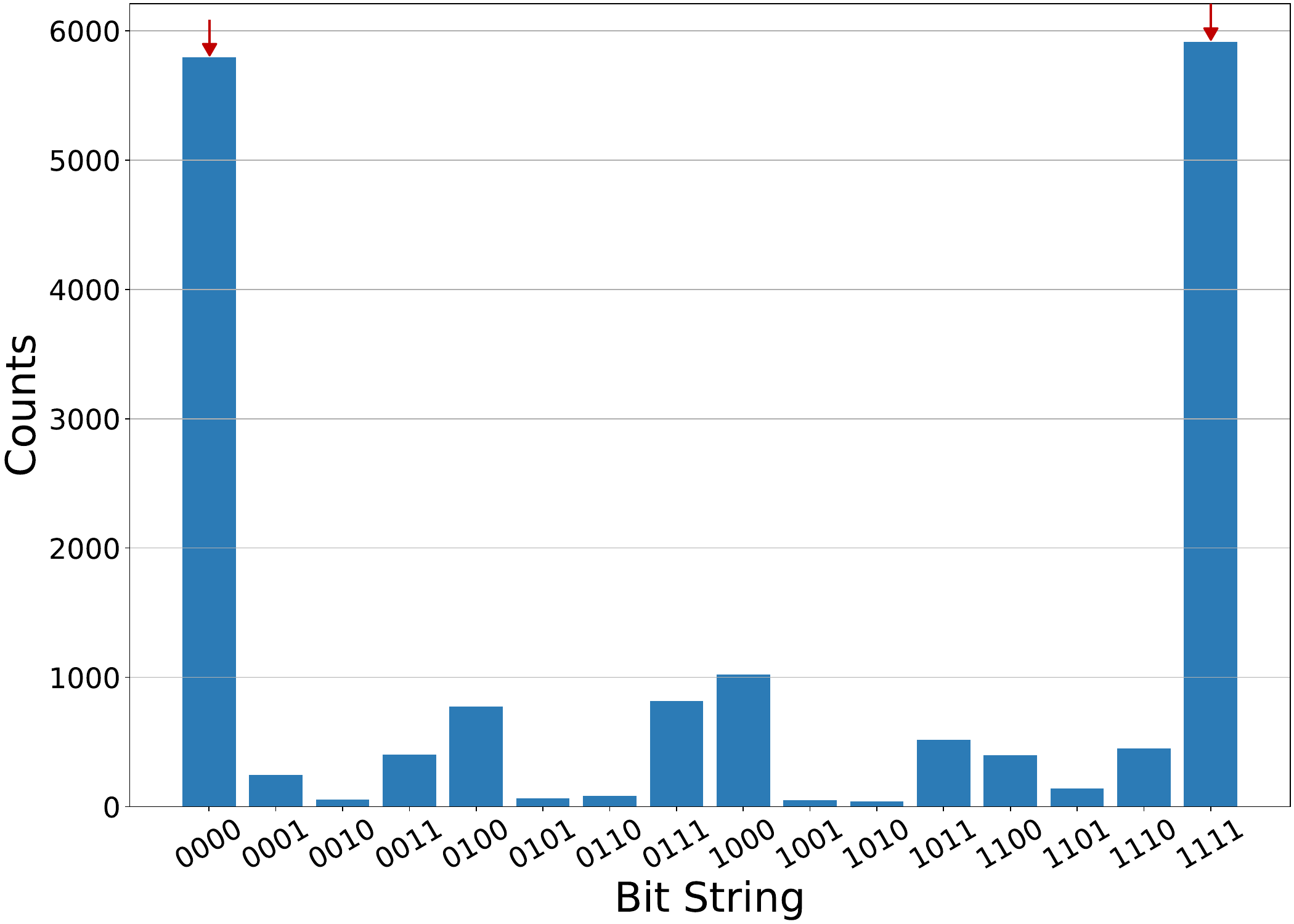}
        \caption{}
    \end{subfigure}
    \begin{subfigure}{0.45\textwidth}
        \includegraphics[scale=0.2]{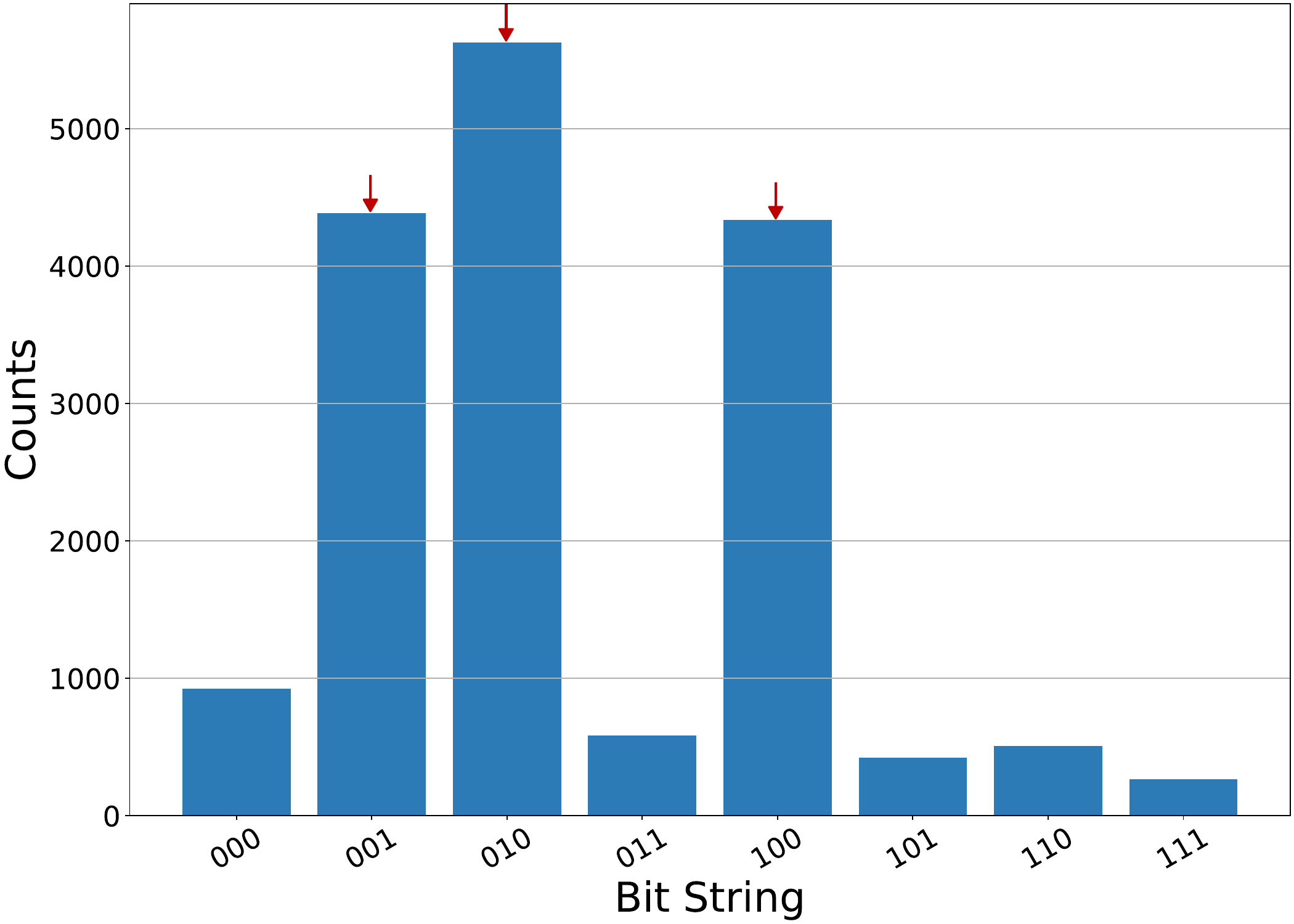}
        \caption{}
    \end{subfigure}
    \caption{Output distribution of the state preparation circuits for (a) $\mathrm{GHZ}_{4}$ and (b) $\mathrm{W}_3$ using dynamic circuits with one ancilla qubit, executed on the AQT superconducting quantum hardware. The correct outputs are highlighted with red arrows.}
        \label{fig:aqt}
\end{figure*}

We first executed two small dynamic circuits—GHZ4 (a 5-qubit circuit including one ancilla) and $\mathrm{W}_3$ (a 4-qubit circuit including one ancilla)—on the Advanced Quantum Testbed (AQT) superconducting quantum machine \cite{aqt} to validate the correctness of our dynamic circuit protocol on hardware. The results of the output distributions are shown in Figure \ref{fig:aqt}, where the correct output states are highlighted with red arrows. Despite the presence of gate and MCM errors, the hardware results confirm that our protocol can successfully generate dynamic circuits for state preparation. 

Then, we execute full dataset of the dynamic circuits and unitary circuits on Qiskit noise simulator {\it Fake127QPulseV1} and IBM quantum hardware $ibm\_nazca$. For hardware execution, we repeat the experiments three times with different qubit mapping allocations. For unitary circuits, we select the best circuits among all the unitary circuit preparation methods. To further reduce the idling errors, we apply SDD. We compare the noisy simulation and hardware results with the ideal simulation results (the ideal simulation results for these state preparation are already known) using the Total Variation Distance (TVD) as the metric for state fidelity validation. The TVD measures the distance between two
probability distributions $P$ and $Q$ over the same sample space, and
is defined as:

\begin{equation}
    \text{TVD}(P, Q) = \frac{1}{2}\sum_x|P(x) - Q(x)|
\end{equation}

\begin{figure*}
    \centering
    \begin{subfigure}{0.45\textwidth}
      \centering
    \includegraphics[scale=0.3]{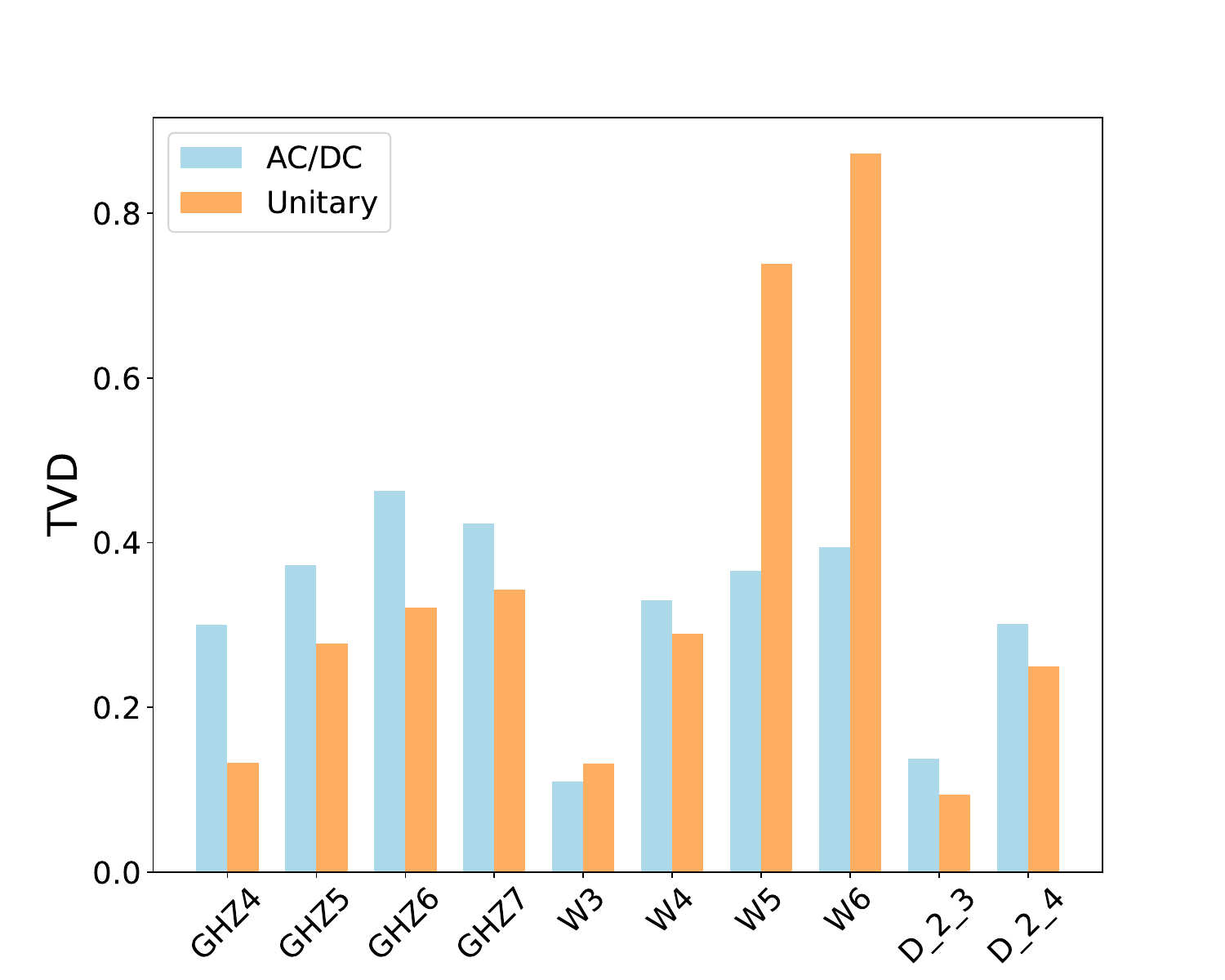}
    \caption{}
    \end{subfigure}
    \begin{subfigure}{0.45\textwidth}
        \includegraphics[scale=0.3]{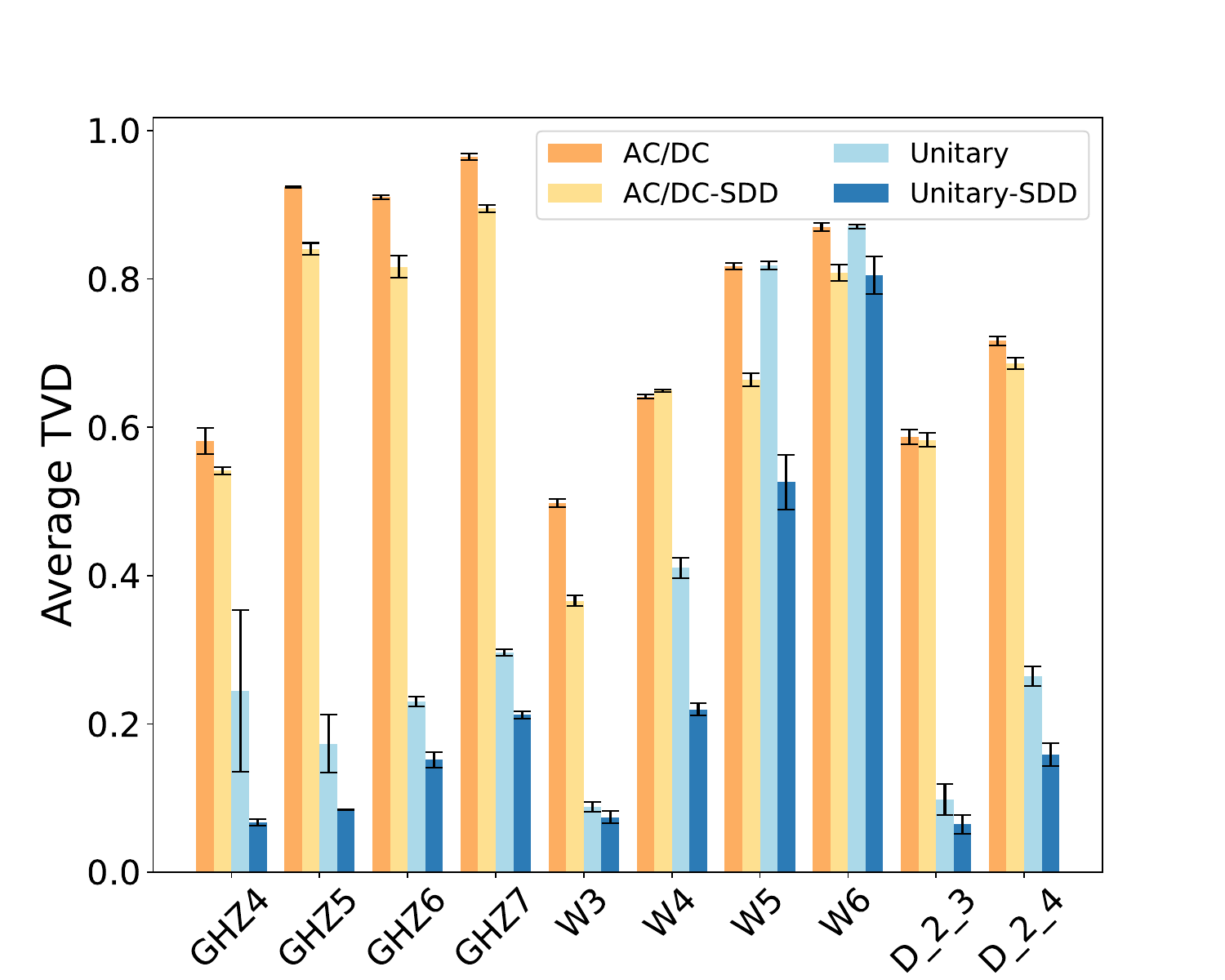}
        \caption{}
    \end{subfigure}
    \caption{The TVD results for executing unitary and dynamic circuits (AC/DC) to prepare various quantum states on (a) noise simulator (b) IBM quantum hardware. SDD represents staggered dynamical decoupling.}
    \label{fig:state-res}
\end{figure*}

The results are shown in Figure~\ref{fig:state-res}. Even though the dynamic circuit can reduce the circuit depth, it often slightly increases the number of CNOTs. For small-scale circuits executed on quantum hardware, circuit depth is not yet a significant limitation. The noisy simulation results show that unitary circuits in general have better state fidelity with smaller TVD, except for some W states, where dynamic circuits give higher fidelity. However, the hardware results show that including just one mid-circuit measurement dramatically reduces circuit fidelity. This reduction in fidelity is due to the fact that mid-circuit measurements have much higher error rates and longer durations compared to other gates on IBM quantum hardware. This observation is consistent with our findings for dynamic circuits used in unitary preparation, which will be discussed in Section~\ref{sec:utryres}.

\subsection{Evaluating the Fidelity Tradeoff}

While our protocol is able to find substantially shorter circuits and, in some cases, fewer gates, the hardware results show that the resultant fidelity is actually worse than that of the unitary circuit described in Section~\ref{sec:state-fid} and Section~\ref{sec:unitary-fid}. This is a symptom of poor fidelity of mid-circuit measurement. As mentioned, these fidelities are expected to improve as hardware progresses to support error correction which requires mid-circuit measurement and feed-forward operations. These advancements will, in turn, boosts the performance of our dynamic circuits. Our goal is to understand: ``How much does  mid-circuit measurement need to improve before our dynamic circuits can outperform unitary circuits?".

In order to perform this analysis, we leverage the framework introduced in \cite{kalloor2024roofline}. This tool allows us analyze our compiled circuit parameters (depth, gate count, measurements) across a range of hardware parameters (qubit coherence, gate fidelities, gate times, etc.). To evaluate the impact of our depth reduction, we adapt a fidelity model based on \cite{tan_qubit_mapping, tan2024compilation} to our case:
\begin{align}
    \mathbf{F} = f_2^{n_2} \cdot f_1^{n_1} \cdot (f_{\text{MCM}})^{N \cdot M} \cdot \exp\left(-N \cdot \frac{C_T}{T_1}\right)
\end{align}
$N$ is the number of qubits, $T_1$ is the qubit coherence time, $C_T$ is the time needed to run a circuit, $f_i$ is the gate fidelity of an $i$-qubit gate, $n_i$ is the number of $i$-qubit gates, $f_{\text{MCM}}$ is the MCM fidelity, $M$ is the number of mid-circuit measurement cycles, i.e., simultaneous measurement is considered as one cycle. In this model, we assume that a mid-circuit measurement lowers the fidelity of all qubits (including idle qubits), due to measurement crosstalk. Thus, we have an additional factor of $N$ in the third term. The circuit time is a sum of the gate times along the critical path. For dynamic circuits, there is an additional feed-forward time ($t_{\text{FF}})$ that we include which factors in the time it takes to pass the control signal. 

With this model, we are able to define our objective function ($\Pi$) as the difference in fidelity between the original unitary circuit and our dynamic circuit:
\begin{align}
  \Pi = \mathbf{F}_{\text{unitary}} - \mathbf{F}_{\text{dynamic}}
\end{align}
We want to understand at what fidelity ($f_{\text{MCM}}$) does our objective function become 0. This value is the desired mid-circuit measurement fidelity we require to outperform a unitary circuit.

\begin{figure*}[htbp]
    \centering
    \begin{subfigure}[b]{0.45\textwidth}
        \centering
        \includegraphics[scale=0.30]{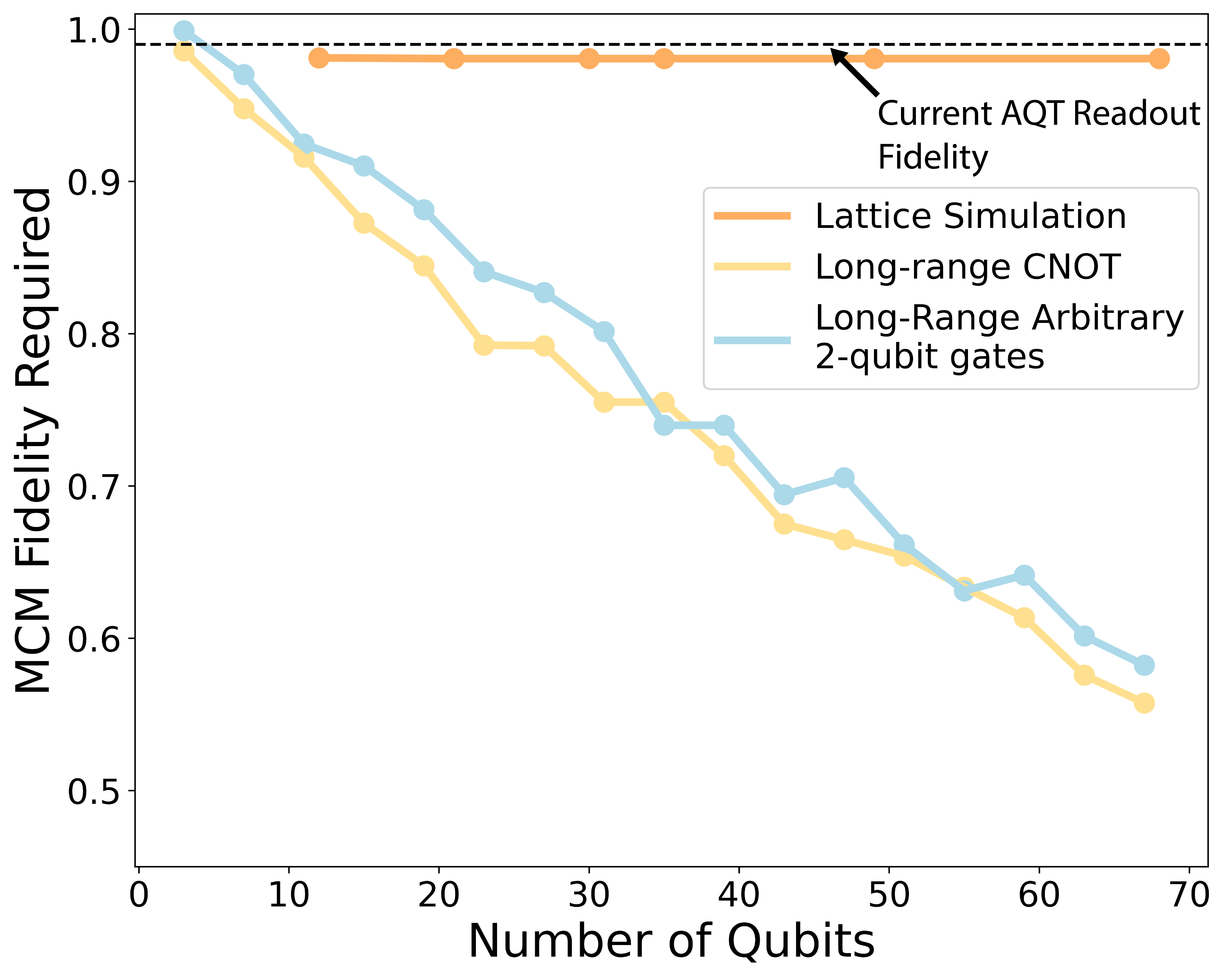}
        \caption{}
    \end{subfigure}
    \begin{subfigure}[b]{0.45\textwidth}
        \centering
        \includegraphics[scale=0.30]{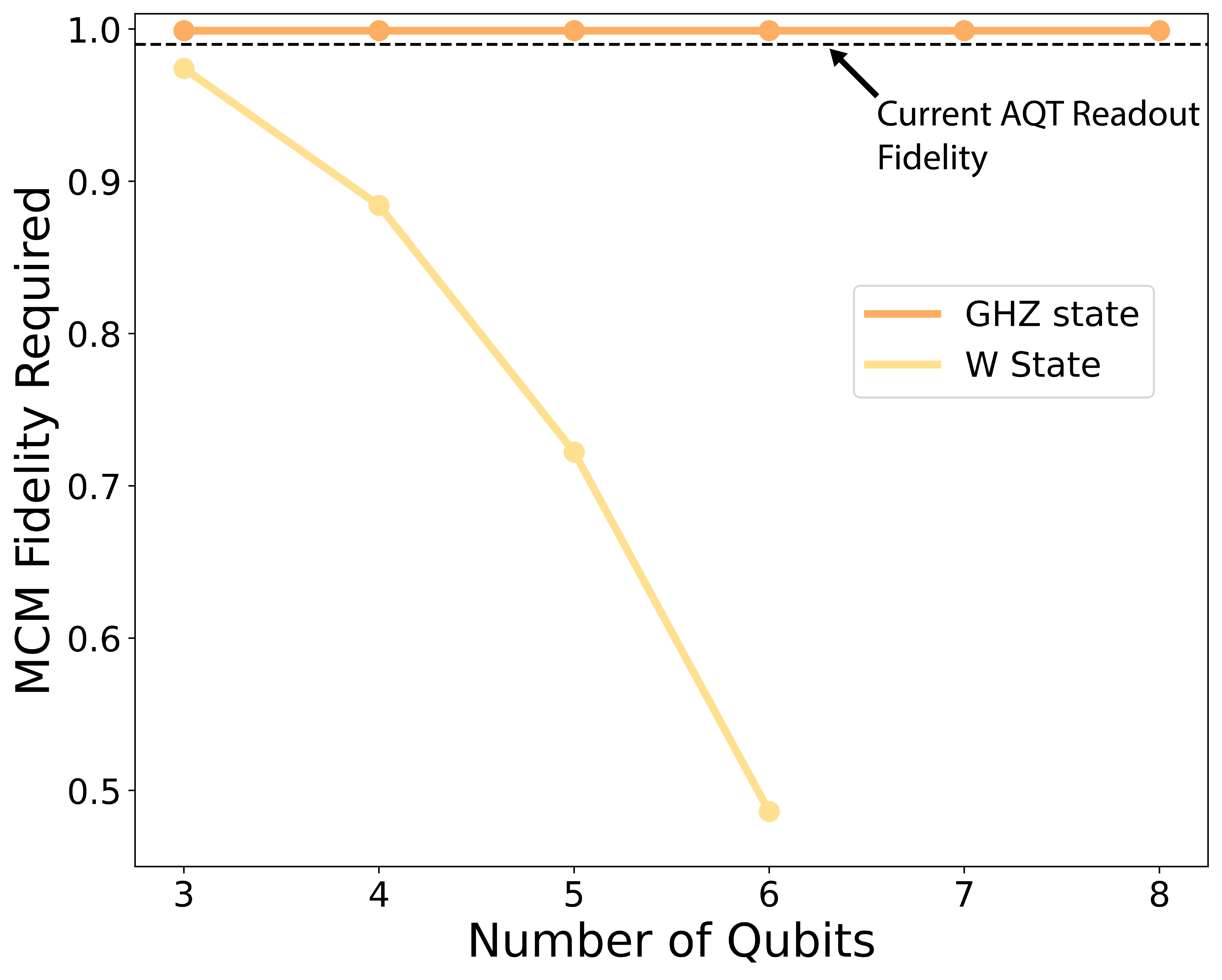}
        \label{fig:sub1}
        \caption{}
    \end{subfigure}
    \caption{The mid-circuit measurement (MCM) fidelity required for dynamic circuits to outperform unitary circuits across our benchmarks for (a) unitary preparation and (b) state preparation for different circuit widths. We show the current readout fidelity of AQT superconducting quantum computer for baseline as we expect MCM fidelities to improve to match readout fidelities. The dramatic decrease of required MCM fidelity is due to the reduction of circuit depth.
    }
    \label{fig:dyn_roofline}
\end{figure*}

Our hardware parameters are derived from measured benchmarks on the AQT superconducting machine \cite{aqt_benchmarks}. The qubit coherence ($T_1$) is $86.6 ~\mu$s. The gate fidelities ($f_1, f_2$) of the one and two-qubit gates are $0.999$ and $0.99$ respectively, while the gate times are $60~n$s and $400~n$s. The feed-forward time ($t_{\text{FF}}$) is $150~n$s. While we vary the $f_{\text{MCM}}$ in our models, we plot the current average readout fidelity ($f_{M}=0.989$) of AQT hardware as a reasonable estimate for future mid-circuit measurement fidelities. Based on these parameters, we obtain the results shown in Figure \ref{fig:dyn_roofline}. Note that the noise analysis results we obtain are strongly dependent on the noise model, hardware parameters, and target quantum platform. A similar noise analysis can be performed for other quantum platforms.

As shown in Figure \ref{fig:dyn_roofline} (a), in unitary preparation for long-range 2-qubit gates, the circuit depth remains constant for the dynamic circuit but increases with the number of qubits for the unitary circuit, significantly reducing the MCM fidelity requirement as the circuit width increases. For lattice simulation, although dynamic circuits introduce many MCMs and we assume that they have impact on all the qubits, the reduction in circuit depth and gate count leads to a lower MCM fidelity requirement compared to the current average AQT qubit readout fidelity. Note that this is a worst-case assumption, as MCMs are more likely to impact neighboring (spectator) qubits rather than all qubits. Therefore, in reality, the fidelity requirement is expected to be lower. Moreover, since high MCM fidelity is ultimately needed to achieve QEC, we anticipate that dynamic circuits will outperform unitary circuits in the future. 

As shown in Figure \ref{fig:dyn_roofline} (a), for state preparation, introducing one ancilla qubit using dynamic circuits to prepare a GHZ states consistently requires very high MCM fidelity to outperform the unitary circuit version. This is because adding an ancilla does not significantly reduce the circuit depth compared to the efficient GHZ state preparation method using unitary circuits, as discussed in Section~\ref{sec:ghz}. In contrast, for W state preparation, the MCM fidelity requirement for dynamic circuits decreases as the number of qubits increases, due to the significant circuit depth reduction achieved through dynamic circuits, as elaborated in Section~\ref{sec:w}.
\section{Discussion and Conclusion}

In our paper, we present AC/DC, an automated mathematical framework for generating dynamic circuits to prepare arbitrary states and unitaries. Our protocol has been validated on simulator and quantum hardware. Additionally, we demonstrate a physical application in lattice simulations for spin models using dynamic circuits, showing potential reductions in circuit depth and gate count for large quantum circuits. Our prototype can be fine-tuned to build dynamic circuits for specific quantum algorithms, thereby optimizing quantum resource usage.

Dynamic quantum circuits are an essential primitive for the future of quantum computing. Not only can they be leveraged to reduce circuit depth, but as the core of any QEC code implementation, we hope that our proposed dynamic circuit protocol will facilitate the discovery of new methods for magic state injection and distillation. Our hardware results indicate a clear trade-off between resource reduction achieved through dynamic circuits and the increased error rates introduced by mid-circuit measurements and feed-forward loops. This trade-off must be considered when implementing algorithm-specific dynamic circuits. Given that the hardware engineering of mid-circuit measurements is expected to improve in the future, particularly in the context of fault-tolerant quantum computing, the potential of dynamic circuits is immense.

For simplicity, we allocate a layer of $U_3$ gates for each branch circuit. Theoretically, there is more flexibility in designing branch circuits with unitaries with both single qubit gates and entangled gates,  which might further reduce the circuit depth. Synthesizing multiple unitaries simultaneously—including the base unitary $U_b$ applied to all qubits and the branch circuit unitaries applied only to system qubits—is challenging and time-consuming. Designing a better synthesis algorithm that leverages the additional freedom in branch circuit unitaries is an area for future work. Our cost function can be easily adapted to accommodate this.

\begin{acks}
This work was supported by the U.S. Department of Energy, Office of Science, Office of Advanced Scientific Computing Research through the Accelerated Research in Quantum Computing Program MACH-Q Project. This research used resources of the Oak Ridge Leadership Computing Facility, which is a DOE Office of Science User Facility supported under Contract No.~DE-AC05-00OR22725. This research used resources of the National Energy Research Scientific Computing Center (NERSC), a Department of Energy Office of Science User Facility using NERSC award DDR-ERCAPm4141. A.H.~acknowledges financial support from the U.S.~Department of Energy, Office of Science, Office of Advanced Scientific Computing Research Quantum Testbed Program under Contract No.~DE-AC02-05CH11231.
\end{acks}

\bibliographystyle{ACM-Reference-Format}
\bibliography{sample-base}

\appendix

\section{Cost function for Unitary Preparation}

\label{appdx:protocol-unitary}

Suppose we have $s$ system qubits
and $a$ ancilla qubits, and we aim to prepare the target unitary
$V$ on the system qubits. Initially, all the ancilla
qubits are in $|0\rangle$ state. Because we can defer the measurement, without loss of generality, we can consider the following: we apply a unitary $U$ to  both the system and the ancilla qubits, and measure the ancilla qubits at the end. In order for this to represent a unitary $V$ on the system qubits, the following should hold for every system state $\ket{\psi}$ up to a global phase:
\begin{align}\label{eq:unitary_prep_goal}
    U (\ket{\psi} \otimes \ket{0^{\otimes a}}) = e^{i\phi} (V\ket{\psi}) \otimes \ket{\alpha}.
\end{align}
Here $\ket{\alpha}$ is a generic state for the ancillas, and system qubits and ancilla qubits are completely disentangled after we apply $U$. Because the ancilla qubits are measured at the end, as long as they are disentangled from the system qubit, the state $\ket{\alpha}$ can be any $a$-qubit state. As we had in Appendix \ref{appx:state}, both $U$ and $\ket{\alpha}$ are parametrized, and we will vary them until the equality holds. We can define $ \ket{\alpha} = W \ket{0^{\otimes a}}$, and rewrite the equation as the following
\begin{align}\label{eq:unitary_prep_goal}
    (U - e^{i\phi} V \otimes W )  \ket{\psi} \otimes \ket{0^{\otimes a}} = 0.
\end{align}
%
In this section, we will construct a cost function for this problem so that we can find a solution for this unitary $U$ variationally. To do so, we will separate the problem into two pieces. The first part is constructing a faithful cost function $C_{\text{sub}}(U)$ to find a unitary $U$ that is equal to the target unitary only in a subspace. In our case, this subspace consists of $\ket{\psi} \otimes \ket{0^{\otimes a}}$, i.e. the states that have all ancilla qubits in the $\ket{0}$ state. The second part is to modify $C_{\text{sub}}(U)$ to our specific goal given in Eq.~\eqref{eq:unitary_prep_goal}. This will lead to the dynamic circuit cost function of type 2, $C_{\text{dyn1}}(U,W)$, given in Eq. \eqref{eq:Cdyn1-unitary}. Furthermore, we will optimize the arbitrary unitary $W$ for a given $U$, and obtain the dynamic circuit cost function of type 2, $C_{\text{dyn2}}(U)$, given in Eq. \eqref{eq:Cdyn2}.

\subsection{Proof for $C_\mathrm{sub}$}
\label{appx:Csub}
We would like to find a unitary $U$ that is equal to a target unitary $U_T$ up to a global phase, on the subspace containing states $\ket{\psi} \otimes \ket{0^{\otimes a}}$ that has the ancilla qubits in the $\ket{0}$ state. Thus, our goal is to obtain $U$ such that
\begin{align}\label{eq:sub_space_goal}
    (U - e^{i\phi} U_T) \ket{\psi} \otimes \ket{0^{\otimes a}} = 0,
\end{align}
for all $s$-qubit states $\ket{\psi}$. In this subsection, we will construct a faithful cost function $C_{\mathrm{sub}}(U)$ such that minimizing $C_{\mathrm{sub}}(U)$ is equivalent to solving Eq. \eqref{eq:sub_space_goal}.

Because Eq.~\eqref{eq:sub_space_goal} is a linear equation, it is sufficient if it is satisfied for all computational basis states $\ket{i} \otimes \ket{0^{\otimes a}}$ for $i = 0,1,\dots,2^s-1$. Defining $A := U - e^{i\phi} U_T$ for convenience, we see that for each computational basis, we would like to minimize the following quantity:
\begin{align}
\begin{split}
    C_i(U,\phi) =& \frac{1}{2} \Big|\Big| (U - e^{i\phi} U_T) \ket{i \otimes 0^{\otimes a}} \Big|\Big|^2 = \frac{1}{2} \Big|\Big| A \ket{i \otimes 0^{\otimes a}} \Big|\Big|^2 = \frac{1}{2}\bra{i \otimes 0^{\otimes a}} A^\dagger A \ket{i \otimes 0^{\otimes a}} , 
\end{split}
\end{align}
%
The factor of $1/2$ is included to ensure $0 \leq C_i(U.\phi) \leq 1$.

Intuitively, we can propose the average of $C_i$ as the cost function for our problem:
\begin{align}\label{eq:Csub_compact}
\begin{split}
    C_{\text{sub}}(U, \phi) = \frac{1}{2^s} \sum_{i=0}^{2^s-1}C_i(U,\phi) = \frac{1}{2^{s+1}} \sum_{i=0}^{2^s-1} \Tr \left( \ket{i \otimes 0^{\otimes a}} \bra{i \otimes 0^{\otimes a}} A^\dagger A \right) = \frac{1}{2^{s+1}}  \Tr \left( f_0 A^\dagger A \right),
\end{split}
\end{align}
where we define $f_0 := I^{\otimes s} \otimes (\ket{0}\bra{0})^{\otimes a}$ as also mentioned below Eq. \eqref{eq:Cdyn1-unitary}.
We will prove the faithfulness of $C_{\text{sub}}$ by using this compact form. Note that because $0\leq C_i\leq1$, then their average also is between $0$ and $1$, which yields $0\leq C_{\text{sub}}\leq 1$. We know that if Eq. \eqref{eq:sub_space_goal} is satisfied, then $C_{\text{sub}} = 0$. The reverse is also true. If $C_{\text{sub}} = 0$, we obtain $\Tr \left( f_0 A^\dagger A \right) = 0$. Using $f_0^2 = f_0$ and $f^\dagger_0 = f_0$, this would mean that $\Tr((A f_0)^\dagger (A f_0)) = 0$. Because $(A f_0)^\dagger (A f_0)$ is a positive semi-definite matrix, this yields $A f_0 = 0$. Applying this matrix to a state $\ket{\psi} \otimes \ket{0}^{\otimes a}$, we obtain $0 = A f_0 \ket{\psi} \otimes \ket{0}^{\otimes a} = A \ket{\psi} \otimes \ket{0}^{\otimes a}$, which is Eq. \eqref{eq:sub_space_goal}. Thus, $C_{\text{sub}} = 0$ if and only if $A \ket{\psi} \otimes \ket{0}^{\otimes a} = 0$, which proves its faithfulness.

We can optimize the angle $\phi$ for a given $U$, and obtain an equivalent cost function with fewer parameters. Plugging in $A = U - e^{i\phi} U_T$ in Eq. \eqref{eq:Csub_compact}, we obtain
\begin{align}\label{eq:Csub_simplifiying}
\begin{split}
    C_{\text{sub}}(U, \phi) &= \frac{1}{2^{s+1}} \Tr \left( f_0 (U - e^{i\phi} U_T)^\dagger (U - e^{i\phi} U_T) \right) \\
    &= \frac{1}{2^{s+1}} \Tr \left( f_0 (2 - e^{i\phi}U^\dagger U_T - e^{-i\phi} U^\dagger_T U) \right) \\
    &= 1 - \frac{1}{2^{s}} \Re \Big( e^{-i\phi} \:\Tr \left(f_0 U^\dagger_T U \right) \Big). 
\end{split}
\end{align}
Choosing the optimum $\phi$ for a given $U$ and $U_T$ will replace the real part with an absolute value. Then we would obtain the following cost function 
\begin{align}\label{eq:Csub_simple}
    C_{\text{sub}}(U) &= 1 - \frac{1}{2^{s}} \left|\Tr \left(f_0 U^\dagger_T U \right) \right|.
\end{align}
In the following subsection, we will use this cost function to construct the dynamic circuit cost functions of type 1 and type 2.

\subsection{Dynamic circuit cost functions of type 1 and 2}
\label{appx:Csub2}

In this section, we will generate two cost functions to solve Eq. \eqref{eq:unitary_prep_goal}. This equation can be obtained from Eq. \eqref{eq:sub_space_goal} by replacing $U \leftarrow U$ and $U_T \leftarrow V \otimes W$. Plugging these in Eq. \eqref{eq:Csub_simple}, we obtain the following as a cost function:
\begin{align}\label{aeq:Cdyn1}
    C_{\mathrm{dyn1}}(U,W) = 1 - \frac{1}{2^s} \left | \Tr(f_0 (V \otimes W)^\dagger \:U) \right |.
\end{align}
This is the cost function we have provided in Eq. \eqref{eq:Cdyn1-unitary}. As it can be seen, because it is derived from a faithful cost function for a subspace unitary problem, this cost function is faithful for solving the unitary preparation problem Eq. \eqref{eq:unitary_prep_goal}.

We can further optimize the $W$ for a given $U$ and $V$, and 
and therefore obtain the dynamic circuit cost function of type 2, which we provide in Eq. \eqref{eq:Cdyn2}. To do so, 
let us write $f_0$ explicitly and plug in $W\ket{0^{\otimes a}} = \ket{\alpha} = \sum_{i=0}^{2^a-1} \alpha_i \ket{i}$. Then we obtain
\begin{align}
\begin{split}
    C_{\mathrm{dyn1}}(U,W) &= 1 - \frac{1}{2^s} \left | \Tr((I \otimes \ket{0^{\otimes a}} \bra{0^{\otimes a}}) (V \otimes W)^\dagger \:U) \right | \\
    &= 1 - \frac{1}{2^s} \left | \Tr((V^\dagger \otimes \ket{0^{\otimes a}} \bra{\alpha}) \:U) \right | \\
    &= 1 - \frac{1}{2^s} \left | \sum_{i=0}^{2^a-1} \alpha_i^* \Tr((V^\dagger \otimes \ket{0^{\otimes a}} \bra{i}) \:U) \right |.
\end{split}
\end{align}
Optimizing $W$ is now equivalent to optimizing $\alpha_i^*$. Note that because $W$ is unitary, $\alpha_i$ satisfy the normalization constraint $\sum_i \alpha_i^* \alpha_i = 1$. For convenience, let us define the following
\begin{align}\label{aeq:vecT}
    T_i = \Tr((V^\dagger \otimes \ket{0} \bra{i}) \:U) = \Tr(V^\dagger U_{i,0}),
\end{align}
where we have $U_{i,j} = \Tr_a \left( (I \otimes \bra{i}) \: U \: (I \otimes \ket{j}) \right)$.
Then the cost function simplifies into the following
\begin{align}
\begin{split}
    C_{\mathrm{dyn1}}(U,W) &= 1 - \frac{1}{2^s} \left | \sum_{i=0}^{2^a-1} \alpha_i^* T_i \right |,
\end{split}
\end{align}
which is basically the absolute value of the Euclidean inner product of vectors $\Vec{\alpha}$ and $\Vec{T}$. It is well known that the inner product is always between $-|\Vec{\alpha}|\:|\Vec{T}|$ and $|\Vec{\alpha}|\:|\Vec{T}|$, and the extreme conditions are met when $\Vec{\alpha} = \pm \Vec{T} / |\Vec{T}|$, where $c$ stands for ``critical value". Both of these are a valid solution to our problem, which then leads to
\begin{align}
\begin{split}
    C_{\mathrm{dyn1}}(U,W_c) &= 1 - \frac{1}{2^s} \left | \sum_{i=0}^{2^a-1} (\alpha_i)_c^* T_i \right | \\
    &= 1 - \frac{1}{2^s |\Vec{T}|} \left | \sum_{i=0}^{2^a-1} T_i^* T_i \right | \\
    &= 1 - \frac{|\vec{T}|}{2^s}.
\end{split}
\end{align}
Thus, minimizing $C_{\mathrm{dyn1}}(U,W)$ is equivalent to minimizing $C_{\mathrm{dyn1}}(U,W_c)$, which is equivalent to maximizing the norm of the vector $\vec{T}$. Therefore, $C_{\mathrm{dyn1}}(U,W_c)$ is equivalent to the following cost function
\begin{align}\label{aeq:Cdyn2}
\begin{split}
    C_{\mathrm{dyn2}}(U) &= 1 - \frac{|\vec{T}|^2}{4^s},
\end{split}
\end{align}
because both are aiming to maximize $|\vec{T}|$, both are equal to $0$ if and only if we have $U = V \otimes I^{\otimes a}$ for states $\ket{\psi} \otimes \ket{0^{\otimes a}}$, and both are normalized to be bounded between $0$ and $1$. Plugging in Eq. \eqref{aeq:vecT}, we then obtain 
\begin{align}
\begin{split}
    C_{\mathrm{dyn2}}(U) &= 1 - \frac{1}{4^s} \sum_{i=0}^{2^a-1} \left|\Tr(V^\dagger U_{i,0}) \right|^2,
\end{split}
\end{align}
which is precisely the cost function we have introduced in Eq.~\eqref{eq:Cdyn2}. As for $C_{\mathrm{dyn1}}(U,W)$, the derivation of $C_{\mathrm{dyn2}}(U)$ from a faithful cost function proves its faithfulness for the equation Eq. \eqref{eq:unitary_prep_goal}.

\section{Open quantum system perspective}
\label{appx:oqs}
Dynamic quantum circuits can be viewed as open quantum systems, where the interaction with an ancillary system, acting as an environment, plays a crucial role in their time evolution. In this context, the formalism of quantum channels provides a natural framework for describing the behavior of dynamic quantum circuits, allowing us to analyze their properties and derive a cost function for optimization.

For convenience, we consider the dynamic circuit with simultaneous measurement given in Eq.~\eqref{eq:u_simultaneous}, where $U$ is consisting of a joint system-ancilla unitary $U_b$, followed by a simultaneous measurement of the ancillas and conditional unitaries applied to the system based on the the measurement outcome. This dynamic circuit ansatz attempts to implement a target unitary $V$ on the system.

Treating the dynamic circuit as a open quantum system, we use the formalism of quantum channels. The quantum channel for the dynamic circuit, $\mathcal{D}$, can be written in operator-sum or Kraus representation as~\cite{nielsen2010quantum, wilde2013quantum}
\begin{equation}
  \mathcal{D}(\rho)
  = \sum_{j} K_{j} \rho K_{j}^\dagger,
\end{equation}
where $K_{j}$ is the Kraus operator corresponding to the action on the system, conditioned on measuring the ancilla in the state $|j\rangle$, and application of the branch unitary $U_j$, that is, 
\begin{equation}
    K_{j}
    = U_j \left(
    I^{\otimes s} \otimes \langle j |
    \right) U_b \left(
    I^{\otimes s} \otimes | 0 \rangle^{\otimes a}
    \right),
\end{equation}
and $K_{j}^\dagger$ is
\begin{equation}
    K_{j}^\dagger
    = \left(
    I^{\otimes s} \otimes \langle 0 |^{\otimes a}
    \right) U_b^\dagger \left(
    I^{\otimes s} \otimes | j \rangle 
    \right) U_j^\dagger,
\end{equation}
where $j \in \{0, 1, \cdots, 2^a -1\}$ the state of the ancilla after the measurement $|j\rangle$ and the branch unitary $U_j$. Note that the Kraus operator $K_j$ is equal to $U_{j,0}$ given in Eq.~\eqref{eq:u_ij}. 

The Kraus operators satisfy the relation $\sum_{j} K_j^\dagger K_j = I^{\otimes s}$ as 
\begin{equation}
\begin{aligned}
    \sum_{j} K_j^\dagger K_j
    &
    = \sum_{j}
    \left(
    I^{\otimes s} \otimes \langle 0 |
    \right) U_b^\dagger \left(
    I^{\otimes s} \otimes | j \rangle 
    \right) U_j^\dagger
     U_j \left(
    I^{\otimes s} \otimes \langle j |
    \right) U_b \left(
    I^{\otimes s} \otimes | 0 \rangle
    \right)
    \\ &
    = \left(
    I^{\otimes s} \otimes \langle 0 |
    \right) U_b^\dagger \left(
    I^{\otimes s} \otimes \sum_{j} | j \rangle \langle j |
    \right) U_b \left(
    I^{\otimes s} \otimes | 0 \rangle
    \right)
    \\ &
    = \left(
    I^{\otimes s} \otimes \langle 0 |
    \right) U_b^\dagger 
    U_b \left(
    I^{\otimes s} \otimes | 0 \rangle
    \right)
    \\ &
    = \left(
    I^{\otimes s} \otimes \langle 0 |
    \right) I^{\otimes(s+a)} \left(
    I^{\otimes s} \otimes | 0 \rangle
    \right)
    \\ &
    = I^{\otimes s} \otimes \langle 0 | I^{\otimes a} | 0 \rangle
    = I^{\otimes s}.
\end{aligned}
\end{equation}

If this channel implements the target unitary $V$, applying the inverse of the target unitary $V^\dagger$ should undo the effect of the channel. The inverse of the target unitary, $V^\dagger$, represented as a channel has the operator-sum representations
\begin{equation}
  \mathcal{F} (\rho)
  = V^\dagger \rho V,
\end{equation}
which, being a unitary channel, has only one Kraus operator, $V^\dagger$.

Therefore, the target unitary is implemented if the composition of the channels $\mathcal{D}$ for the dynamic circuit and $\mathcal{F}$ for the inverse of the target unitary,
\begin{equation}
    \mathcal{E} = \mathcal{F} \circ \mathcal{D},
\end{equation}
is the identity channel, $\mathcal{I}$ which leave every operator invariant, $\mathcal{I}(\rho) = \rho$, $\forall \rho$.

The channel $\mathcal{E}$ has the operator-sum representation
\begin{equation}
    \mathcal{E}(\rho) 
    = V^\dagger \left(\sum_{j} K_j \rho K_j^\dagger \right) V
    = \sum_{j} V^\dagger K_j \rho K_j^\dagger V.
\end{equation}
Writing the superoperator $\mathcal{E}$ as a matrix $\mathcal{E}_{\mathrm{mat}}$ acting on the vector $\rho_{\mathrm{vec}}$, obtained by flattening the matrix $\rho$, we have the vectorized representation
\begin{equation}
    \mathcal{E}_{\mathrm{mat}} \, (\rho_{\mathrm{vec}})
    = \left(\sum_{j} (V^\dagger K_j) \otimes (K_j^\dagger V)^T\right) \,
    \rho_{\mathrm{vec}},
\end{equation}
where $A^T$ denotes the transpose of $A$.

Written as a matrix, the eigenvalues of a quantum channel are complex numbers that lie within the unit disc~\cite{wolf2010inverse, wolf2012quantumchannel}. All eigenvalues of the matrix are $+1$ if and only if the channel is the identity channel~\cite{wolf2010inverse, wolf2012quantumchannel}. Therefore, the trace of the channel is maximum if and only if the channel is identity~\cite{wolf2010inverse, wolf2012quantumchannel}. The dimension of $\mathcal{E}_{\mathrm{mat}}$ is $d^2 \times d^2$, where $d = 2^s$, is the dimension of the system Hilbert space. For an identity channel, $\Tr(\mathcal{I}_{\mathrm{mat}}) = d^2 = 4^s$. Therefore, the function
\begin{equation}
    C = 1 - \frac{1}{4^s} \Tr(\mathcal{E}_{\mathrm{mat}}),
\end{equation}
is an effective cost function which is zero if and only if the channel $\mathcal{E}$ is the identity channel, which corresponds to the dynamic circuit channel $\mathcal{D}$ implementing the target unitary. The trace of $\Tr(\mathcal{E}_{\mathrm{mat}})$ reads
\begin{equation}
\begin{aligned}
    \Tr(\mathcal{E}_{\mathrm{mat}})
    &
    = \Tr \left(
    \sum_{j} (V^\dagger K_j) \otimes (K_j^\dagger V)^T
    \right)
    \\ &
    = \sum_{j}
    \Tr \left(V^\dagger K_j \right) \cdot
    \Tr \left( (K_j^\dagger V)^T\right)
    \\ &
    = \sum_{j} \left| \Tr \left(V^\dagger K_j \right) \right|^2,
\end{aligned}
\end{equation}
giving the cost function
\begin{equation}
    C = 1 - \frac{1}{4^s} \sum_{j} \left| \Tr \left(V^\dagger K_j \right) \right|^2,
\end{equation}
which is same as the cost function $C_{\mathrm{dyn2}}$ in Eq.~\eqref{eq:Cdyn2} with the Kraus operator $K_j = U_{j,0}$ for dynamic circuits with simultaneous measurement. This perspective can be generalized to other types of dynamic circuits with different placement of mid-circuit measurements given in Section~\ref{sec:measure}.

\section{Cost function for state preparation}
\label{appx:state}

Suppose we have $s$ system qubits
and $a$ ancilla qubits, and we aim to prepare the target state
$|T\rangle$ on the system qubits. Initially, all the $s$ and $a$
qubits are in $|0\rangle$ state. We apply a unitary $U$ to  both the system and the ancilla qubits. In order to have the system qubits on the target state, the following equality should hold, up to a global phase:
\begin{align}\label{eq:state_prep_goal}
    U \ket{0^{\otimes s+a}} = \ket{T} \otimes \ket{\alpha} = \ket{T \otimes \alpha}.
\end{align}
Here $\ket{\alpha}$ is a generic state for the ancillas, and system qubits and ancilla qubits are completely disentangled. The unitary $U$ we propose here is the same large unitary we introduced in Section \ref{sec:measure}.

To find such a unitary $U$, we need to maximize the fidelity between the left and right-hand sides of the Eq. \eqref{eq:state_prep_goal}, which leads to the following cost function
\begin{align}
    \Tilde{C}(U) = 1 - |\bra{T \otimes \alpha} U \ket{0^{\otimes s+a}} |.
\end{align}
This cost function is well known to be faithful and meaningful for state preparation, and is normalized to have values between $0$ and $1$. 

For our problem, the state $\ket{\alpha}$ and the unitary $U$ are parametrized, and we will vary them to minimize $\Tilde{C}(U)$. Let us write $\ket{\alpha} = \sum_{i = 0}^{2^a-1} \alpha_i \ket{i}$. Then we obtain
\begin{align}
\begin{split}
    \Tilde{C}(U) &= 1 - \left|\sum_i \alpha^*_i \bra{T \otimes i} U \ket{0^{\otimes s+a}}\right|, \\
    &=  1 - \left|\sum_i \alpha^*_i A_i \right|,
\end{split}
\end{align}
where we introduce $A_i := \bra{T \otimes i} U \ket{0^{\otimes s+a}}$ as a shorthand notation. In this compact form, we see that the cost function is basically the absolute value of the Euclidean inner product of vectors $\Vec{\alpha}$ and $\Vec{A}$. It is well known that the inner product is always between $-|\Vec{\alpha}|\:|\Vec{A}|$ and $|\Vec{\alpha}|\:|\Vec{A}|$, and the extreme conditions are met when $\Vec{\alpha} = \Vec{\alpha}_c = \pm \Vec{A} / |\Vec{A}|$, where $c$ stands for ``critical value". Both of these are a valid solution to our problem, which then leads to
\begin{align}
\begin{split}
    \Tilde{C}(U) &= 1 - \left|\sum_i (\alpha_i)_c^*  A_i \right| \\
    &= 1 - \frac{1}{|\vec{A}|}\left|\sum_i A_i^*  A_i \right| \\
    &= 1 - |\vec{A}|.
\end{split}
\end{align}
Now, we know that this cost function is bounded by $0$ and $1$, and takes the value of $0$ if and only if Eq. \eqref{eq:state_prep_goal} is satisfied. Thus, the following function satisfies the same properties:
\begin{align}
\begin{split}
    C(U) &= 1 - |\vec{A}|^2 \\
    &= 1 - \sum_i A^*_i A_i \\
    & = 1 - \sum_i \left| \bra{T \otimes i} U \ket{0^{\otimes s+a}} \right|^2.
\end{split}
\end{align}
$C(U)$ is more preferable to $\Tilde{C}(U)$ because its gradient is easier to calculate.
When we plug in $U = \sum_i (U_i \otimes \ket{i}\bra{i}) \: U_b$, we obtain
\begin{align}
    C(U) = 1- \sum_{i=0}^{2^a - 1} |\langle T \otimes i|  (U_i \otimes |i\rangle \langle i|)\: U_b|0^{\otimes(s+a)}\rangle|^2. 
\end{align}
which is the cost function we have introduced in Eq. \eqref{eq:state_cost}.

\section{Examples of circuits prepared using our dynamic circuit protocol}
\label{appdx:circuits}

\begin{figure}[htbp!]
    \centering
    \begin{subfigure}{\textwidth}
    \centering
        \includegraphics[scale=0.33]{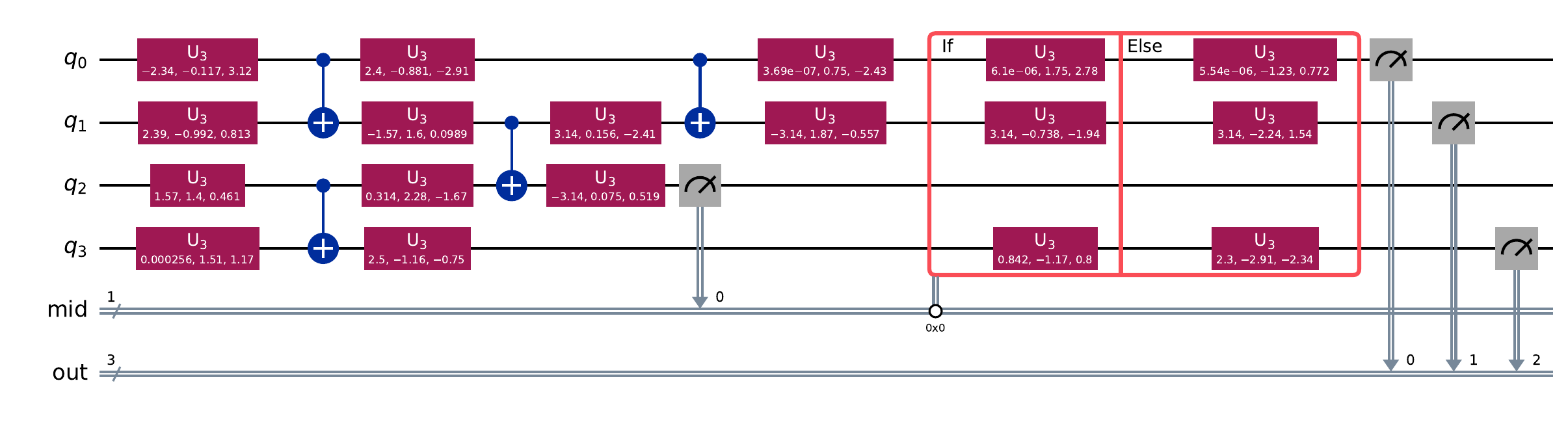}
        \caption{}
    \end{subfigure}
    \begin{subfigure}{\textwidth}
        \includegraphics[scale=0.33]{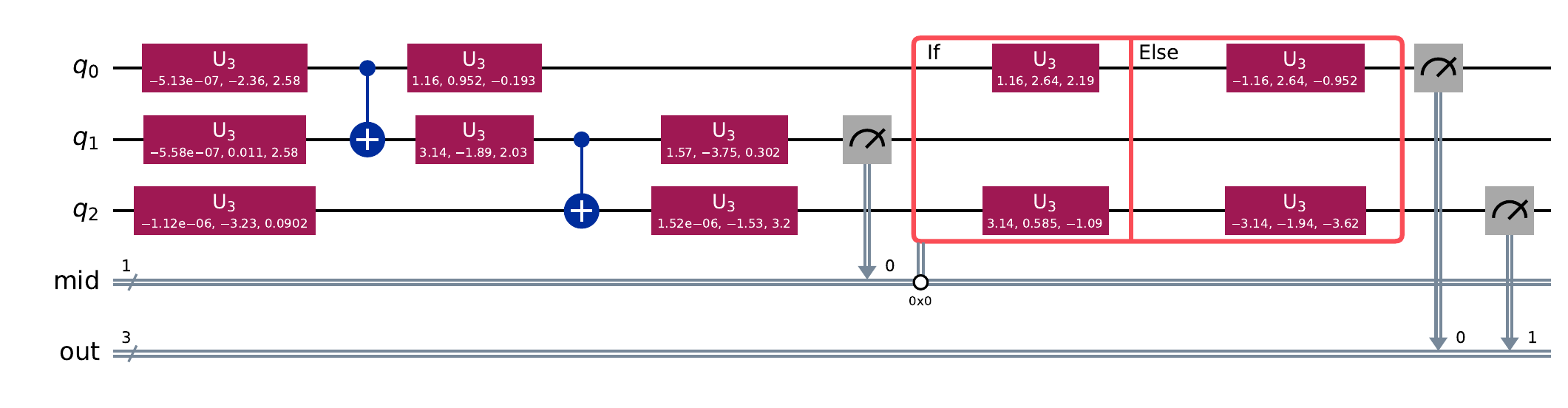}
        \caption{}
    \end{subfigure}

        \begin{subfigure}{\textwidth}
        \includegraphics[scale=0.33]{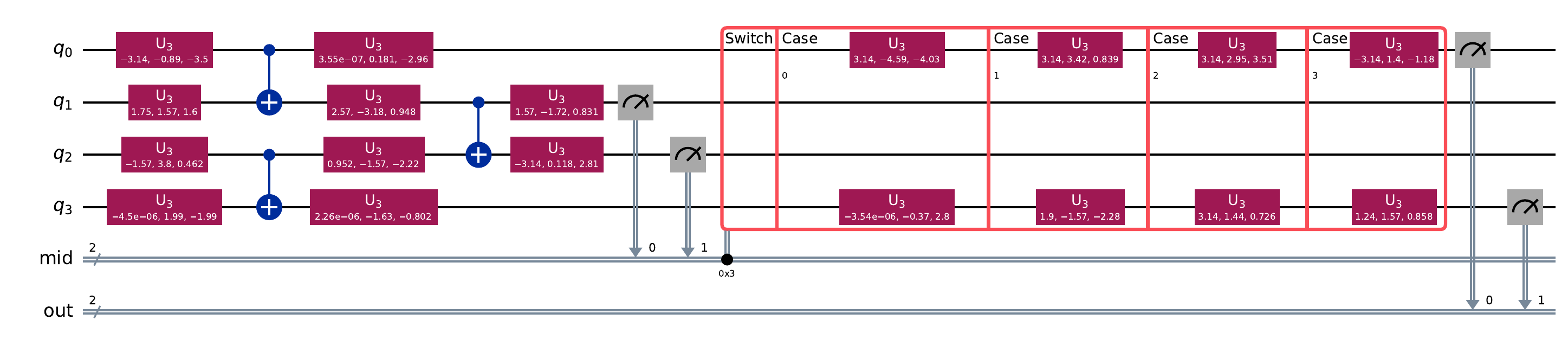}
        \caption{}
    \end{subfigure}
    
        \begin{subfigure}{\textwidth}
        \includegraphics[scale=0.3]{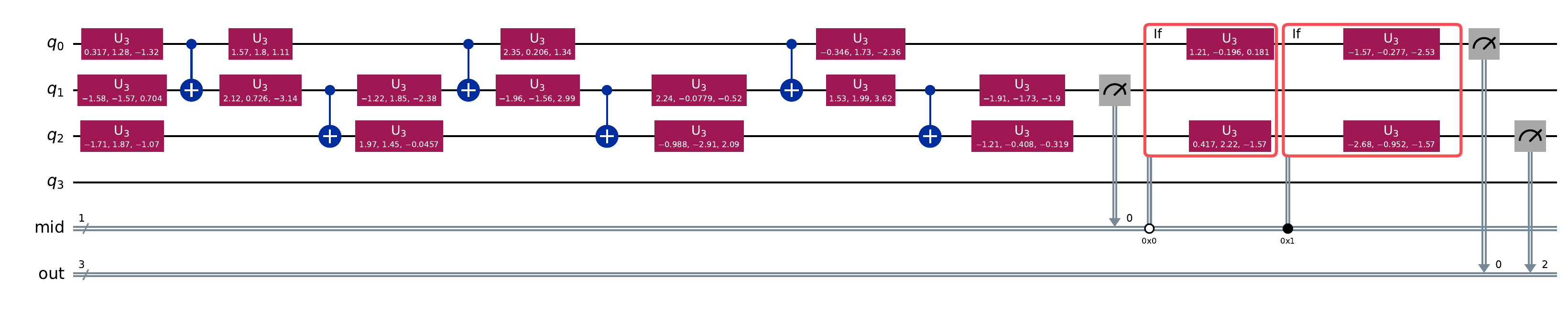}
        \caption{}
        
    \end{subfigure}
        \begin{subfigure}{\textwidth}
        \includegraphics[scale=0.28]{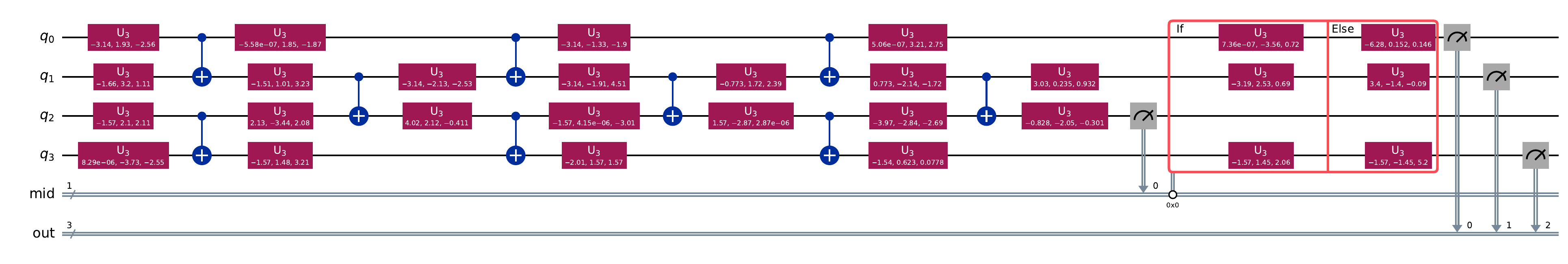}
        \caption{}
    \end{subfigure}
    
    \caption{Dynamic circuits for preparing (a) W3 state on $(q_0, q_1, q_3)$. (b) CNOT between $(q_0, q_2)$. (3) CNOT between $(q_0, q_3)$. (4) Random 2-qubit gate between $(q_0, q_2)$, which corresponds to Rzz($\frac{\pi}{3}$)Rxx($\frac{\pi}{4}$)Ryy($\frac{\pi}{5}$).(4)Toffoli gate on $(q_0, q_1, q_3)$. }
    \label{fig:dc-circuits}
\end{figure}

Figure~\ref{fig:dc-circuits}  illustrates the dynamic circuits generated by our protocol for preparing W states, long-range CNOTs, and the Toffoli gate. Note that the parameters of all the $U_3$ gates are not fixed; they will vary with each instantiation, but the circuit will always be able to prepare the target state or unitary. Moreover, the example circuits shown here follow the basic framework of QSearch, which begins with an initial layer of $U_3$ gates and then blocks are added to the connected qubits. Each block is composed of one CNOT and two $U_3$, like shown in Eq.~\eqref{eq:block}. We can perform gate deletion to remove redundant single-qubit gates for further optimization.

\end{document}